\PassOptionsToPackage{table}{xcolor}
\documentclass{aastex63}
\pdfoutput=1

\usepackage{mathtools}
\usepackage{float}
\usepackage[table]{xcolor}
\usepackage{graphicx,psfrag,amsmath,amsfonts,verbatim}
\usepackage{multirow}
\usepackage{tikz}
\usepackage[caption=false]{subfig}
\usetikzlibrary{arrows.meta}
\tikzset{%
  >={Latex[width=2mm,length=2mm]},
            base/.style = {rectangle, rounded corners, draw=black,
                          minimum width=4cm, minimum height=1cm,
                          text centered, font=\sffamily},
              io/.style = {base, fill=blue!30},
      startstop/.style = {base, fill=red!30},
      decision/.style = {base, fill=green!30},
         process/.style = {base, minimum width=2.5cm, fill=orange!15, font=\ttfamily},
}
\usepackage{url} 
\usepackage{lineno}
\usepackage[inline]{trackchanges} 
\usepackage{soul}
\usepackage{csvsimple,longtable}
\DeclareMathOperator*{\argmin}{argmin} 

\shorttitle{Interpretation of Solar Flare Prediction by LSTM model}
\shortauthors{Sun et al.}

\begin{document}

\title{Interpreting LSTM prediction on Solar Flare Eruption with Time-series Clustering}

\correspondingauthor{Ward Manchester}
\email{chipm@umich.edu}



\author{Hu Sun}
\affiliation{Department of Statistics, University of Michigan\\
Ann Arbor, MI, USA}

\author[0000-0003-0472-9408]{Ward B. Manchester IV}
\affiliation{Department of Climate and Space Sciences and Engineering, University of Michigan \\
Ann Arbor, MI, USA}

\author{Zhenbang Jiao}
\affiliation{Department of Statistics, University of Michigan\\
Ann Arbor, MI, USA}

\author[0000-0002-8963-7432]{Xiantong Wang}
\affiliation{Department of Climate and Space Sciences and Engineering, University of Michigan \\
Ann Arbor, MI, USA}

\author[0000-0002-9516-8134]{Yang Chen}
\affiliation{Department of Statistics, University of Michigan\\
Ann Arbor, MI, USA}

\begin{abstract}
We conduct a post hoc analysis of solar flare predictions made by a Long Short Term Memory (LSTM) model employing data in the form of Space-weather HMI Active Region Patches (SHARP) parameters calculated from data in proximity to the magnetic polarity inversion line where the flares originate.  We train the the LSTM model for binary classification to provide a prediction score for the probability of M/X class flares to occur in next hour. We then develop a dimension-reduction technique to reduce the dimensions of SHARP parameter (LSTM inputs) and demonstrate the different patterns of SHARP parameters corresponding to the transition from low to high prediction score. Our work shows that a subset of SHARP parameters contain the key signals that strong solar flare eruptions are imminent. The dynamics of these parameters have a highly uniform trajectory for many events whose LSTM prediction scores for M/X class flares transition from very low to very high. The results demonstrate the existence of a few threshold values of SHARP parameters that when surpassed indicate a high probability of the eruption of a strong flare. Our method has distilled the knowledge of solar flare eruption learnt by deep learning model and provides a more interpretable approximation, which provides physical insight to processes driving solar flares.
\end{abstract}

%
%

%


%
%
%
%

\section{Introduction}
The Sun exhibits a wide range of eruptive activity, flares, filament eruptions and and coronal mass ejections (CMEs), all of which are magnetically driven (see reviews by \cite{Forbes2000, Schrijver2009, Schmieder2015, Green2018}).  The magnetic free energy driving these eruptions accumulates in the corona in association with electric currents (e.g. \cite{Janvier2014, Schmieder2015, Schmieder2018}). In response to these currents, the non-potential magnetic fields are sheared and twisted as seen in the structure of hot loops observed in the extreme ultraviolet (EUV).  Such EUV images of pre-event corona often show bright loops with sigmoidal structure (e.g. \cite{Canfield1999, Magara2003, Aulanier2010, Green2011}), which are often harbingers of solar eruptions \citep{Falconer2000}. 

In order to study and predict solar eruptions, one would ideally measure the magnetic field where eruptions occur, however the coronal magnetic field is difficult to deduce and impossible to localize beyond the plane of the sky \citep{Dove2011}. In contrast, spectropolarimetry of bright optically-thick spectral lines can provide high-resolution high-cadence reconstruction of the photospheric vector magnetic field in active regions where solar eruptions occur. Examination of magnetic fields has shown signatures strongly associated with solar eruptions that are indicative of  increases in free magnetic energy such as: intensification of the horizontal magnetic fields \citep{Wang2017}, strong magnetic gradients \citep{Schrijver2007}, increases in magnetic shear \citep{Georgoulis2012, Sun2012} increases in magnetic twist \citep{Su2008, Vemareddy2012} and increases in magnetic and current helicities \citep{Tziotziou2012, Wang2018, Vasantharaju2018}.  From these strong associations, predictive parameters and empirical relationships have been derived which are useful for predicting solar flares (e.g. \cite{Falconer2002, Falconer2003, Leka2003a, Leka2003b, Falconer2006, Schrijver2007}). These works tend to focus on flares because they are much more easily observed and categorized by energy level than either filament eruptions or CMEs.  

The application of machine learning is now made possible due to the recent availability of vast quantities of magnetogram data. The most notably source is photospheric vector magnetograms from the Helioseismic and Magnetic Imager (HMI) instrument on the Solar Dynamics Observatory launched in February 2010 \citep{Scherrer2012, Hoeksema2014a}. The first example of machine learning employing these data is that of \cite{Bobra2015}, which was followed by work such as \cite{Liu2017, Nishizuka2017a, Huang2018, Yang2019}, which used  Space-weather HMI Active Region Patches (SHARP) parameters for model training. In their work, \cite{Maranushi2016} and  \cite{Nishizuka2018} developed deep neural networks for flare prediction. A description of these and related work is found in reviews by \cite{Leka2018} and  \cite{Camporeale2019}. 

\begin{figure}[htb]
    \centering
    \includegraphics[width = 0.90\textwidth]{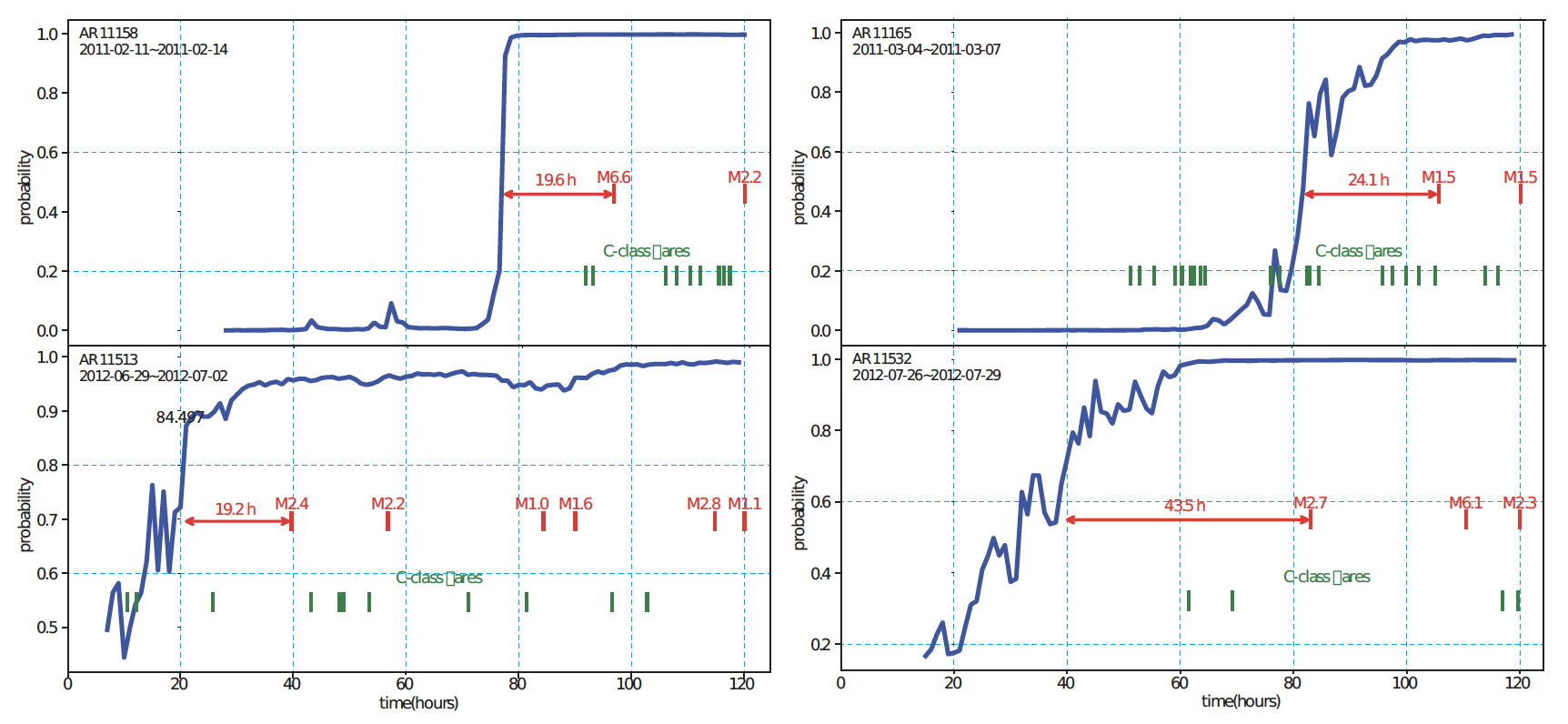}
    \caption{Four examples of prediction scores from the LSTM Strong/weak flare classification model frist presented in \cite{Yang2019}.  Line plots show the probability of an M/X class flare occurring during a 12 hour interval extending into the future for active regions 11158, 11165, 11513, and 11532. Red and green marks indicate the occurrence M/X and class flares respectively.}
    \label{fig:4cases}
\end{figure}

Our work builds upon that of \cite{Yang2019}, which used a Long-Short-Term-Memory (LSTM) model with multi-dimensional time-series SHARP parameter as input to perform binary classification to distinguish strong solar flares of M/X class from weak flares of A/B class. The LSTM model in \cite{Yang2019} outputs a prediction score between 0 and 1 where values close to 1 indicate high probability of an M or X class solar flare. \cite{Yang2019} presented a few case studies (see Figure \ref{fig:4cases}) where the LSTM prediction score (probability of seeing an M/X flare) increases abruptly from nearly 0 to almost unity several hours prior to the event. This rapid change in the flare prediction score shows a strong sensitivity to the input data, which suggests interpretation of the model may provide physical insight on solar flares and conditions necessary for their initiation.   

To answer the question of which SHARP parameter leads to the sudden transition in prediction scores in Figure~\ref{fig:4cases}, \cite{Yang2019} adopted a variable importance measure, which is defined as the testing accuracy when training the LSTM solely on each of the SHARP parameters. However, this does not directly address the problem since the LSTM model involves highly non-linear, non-independent operations on the SHARP parameters, making it extremely hard to disentangle the individual contributions of each SHARP parameter. In this paper, we conduct post-hoc analysis on the fitted LSTM models to obtain interpretability of the deep learning model, thus explaining the intriguing phenomenon of sudden transition in prediction scores shown in Figure~\ref{fig:4cases}.

Specifically, we interpret predictions made by the LSTM model via directly comparing the LSTM model inputs associated with low and high prediction scores. This is achieved via finding a low-dimensional representation of the SHARP parameter (time series) that separates the high prediction score cases from the low prediction score cases. What's challenging is that in the LSTM inputs, we are working with multi-dimensional time series data with both cross-dimensional and temporal dependencies. We combine the dynamic time warping idea with the principle component analysis to project SHARP parameter time series to construct the desired low dimensional spaces. By doing so, we are able to explain the sudden transitions in LSTM prediction scores in Figure~\ref{fig:4cases}, i.e. predicted occurrences of strong flares, in terms of SHARP parameters.

To further improve the interpretability of LSTM model besides analyzing the structure of the inputs, we also refine the input features when training LSTM model. \cite{Yang2019} trained their LSTM model based on SHARP parameters calculated from the full vector field image of the active region. In our paper, we use the SHARP parameters calculated at the neighborhood of the polarity inversion line (PIL), which are more localized to the source of the flare. Selecting data in proximity to PIL filters out noise, reduces irrelevant information, and increases to 35 the number of events showing a sudden probability increase for M/X class events similar to those in Figure \ref{fig:4cases}. However, the weighting of the PIL mask makes the magnitude of all SHARP parameters not directly comparable to the magnitude of the full-image SHARP parameters. 

In the next section, we provide details about the SHARP parameter dataset along the polarity inversion line (PIL), and how we prepare training and test data for LSTM model. Section \ref{LSTMST} revisits the theory of Long-Short-Term-Memory (LSTM) model and also provides the results of LSTM model based on the SHARP parameters along the PIL. In section \ref{LSTMINT}, we outline how we interpret LSTM predictions based on the idea of clustering, and show case studies that fully explain the key SHARP parameters that are closely related to strong solar flare eruption. Section \ref{Conclusion} concludes.

\section{Data Preparation}\label{dataprep}
Our machine learning model undertakes a first-flare classification task, where the positive class is the first M or X flare and the negative class is the first B flare if any. As in \cite{Yang2019}, the predictors used to classify these flares are derived from HMI photospheric vector magnetic field data, which are saved with a cadence of 12 minutes and resolution of 1 arc second \citep{Hoeksema2014a}. From these full-disk data, subsets known as HMI Active Region Patches (HARPs), are produced where data regions are spatially restricted to the near proximity of active regions (ARs). Furthermore, Space-weather HMI Active Region Patches (SHARPs) are an additional data product providing physical parameters calculated from the vector field, which are relevant to solar flare production (see \cite{Bobra2014} for detailed descriptions of these features). In this paper, we use the SHARP parameters that are calculated from vector magnetogram pixels located along the polarity inversion line (PIL). Details about the PIL detection procedure are discussed in \cite{Jingjing2019}. This section introduces the PIL-based SHARP parameter dataset, the steps on how we collect the predictors and the overview of M, X, B first flares for training and testing the machine learning model. The SDO/HMI vector magnetic field images and SHARP parameters are available for download from Joint Science Operations Center (JSOC).

The procedures for locating polarity inversion line in \cite{Jingjing2019} follow the idea of \cite{Schrijver2007}. At any time for a specific HARP region, they took the radial field ($\mathrm{B_r}$) image of the HARP region as input data. Next, they generated two bitmaps for positive and negative pixels, respectively. The positive bitmap labeled all pixels satisfying $\mathrm{B_r>200G}$ as 1 and the negative bitmap labeled all pixels satisfying $\mathrm{B_r<-200G}$ as -1. Then the non-zero pixels in each bitmap are clustered with Density-Based Spatial Clustering of Application with Noise (DBSCAN; \citep{Sander1998}), where positive and negative pixels form clusters that can be considered as positive and negative poles. Finally, they applied Gaussian-Filter convolution to dilate the clusters of strong positive and negative pixels, and then located the polarity inversion line as the intersection area of two dilated clusters of opposite polarity. The PIL mask generated is a weighted mask that puts more weights on pixels who have both strong positive and strong negative pixels in their neighborhood. All SHARP parameters, excluding the ones related to Lorentz force, are calculated following the formulae in Table 3 of \cite{Bobra2014}, and each pixel is weighted by the weight in the PIL mask.

The PIL-based SHARP parameters available are TOTUSJH, TOTPOT, TOTUSJZ, ABSNJZH, SAVNCPP, USFLUX, AREA\_ACR, MEANPOT, R\_VALUE, SHRGT45, MEANSHR, MEANGAM, MEANGBT, MEANGBH, MEANGBZ, MEANJZH, MEANJZD, MEANALP. There are two extra parameters X\_SIZE and Y\_SIZE recording the width and height of each vector field image. The PIL-based SHARP parameters data is of 12-minute cadence. Units of all physical quantities follow that in Table 3 of \cite{Bobra2014}.  Note that the calculation of all physical quantities are weighted by the PIL mask. For example, when the unsigned flux of a pixel is added to the USFLUX of the HARP region at a certain time, the closer the pixel is to the PIL, the larger weight the flux of the pixel will have in the final calculation of USFLUX. All quantities reflect the weighting of the PIL mask. 

We consider events from the NOAA Geostationary Operational Environmental Satellites (GOES) flare list and use corresponding SDO/HMI data from mid 2010 through 2018. Among all the flares, we focus only on the first M-class or X-class flare, whichever comes earlier, and the first B-class flare of each active region, if any. Then we assign each first flare to the corresponding HARP region, based on the mapping between NOAA active region number and HARP region number. We omit those flares whose NOAA active region does not have an associated HARP region number.  Additionally, there are some HARP regions in which B flares happen after its first M/X flare, and those B flares are typically the last flares recorded for that HARP region. We only want to focus on any B flares preceding the first M/X flare, so we discard those B flares sample occurring afterward. Eventually, we have only 681 first flares that have associated PIL-based SHARP parameters, among which there are 163 M/X flares and 518 B flares. For each of the first flares of a certain HARP region, we gather all PIL-based SHARP parameters prior to the flare. Note that the length of the time series of SHARP parameters could vary from region to region since the recording time of different HARP regions prior to its first flare might differ. The PIL-based SHARP parameter data for the first flares are the totality of the data we have prior to data pre-processing.

\begin{figure}[htb]
    \centering
    \includegraphics[width = 0.9\textwidth]{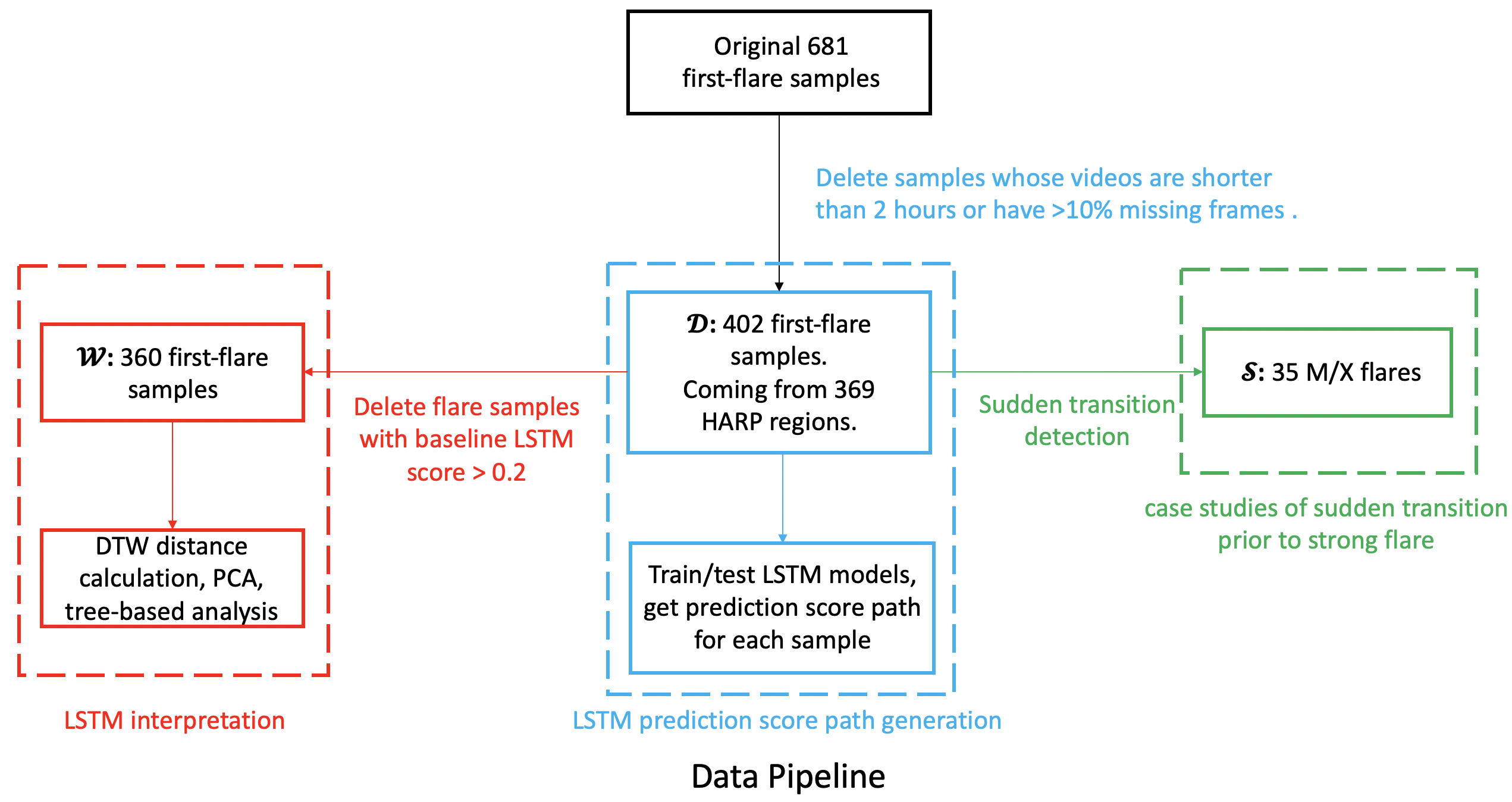}
    \caption{\textbf{Overview of the processed data analyzed in latter sections}. The 681 flare samples are the maximum number of samples we can collect from December 2010 to June 2018 that have the PIL-based SHARP parameters and associated first flares. Data pre-processing leaves us 402 flare samples for LSTM training and testing. We filter 35 M/X flare samples out of the 402 samples that showed a sudden transition of LSTM scores as those in Figure \ref{fig:4cases}. Finally, we select 360 samples out of the 402 samples for LSTM interpretation.}
    \label{fig:dataflow}
\end{figure}

In Figure \ref{fig:dataflow}, we show a simplified pipeline of these 681 first flare events, which we select to 402 samples based on the data pre-processing steps (described in the next subsection) for LSTM training and testing, prediction score path generation and sudden transition detection.  Of these samples, we select 360 samples, whose lowest prediction score is below 0.2, and of these events we find 35 events showing a sudden transition of LSTM scores before their first M/X flares.  To interpret the predictions made by LSTM model, we carry out several steps to determine what the LSTM model has learned about flare eruption. In the remainder of the paper, for simplicity of notations, we refer to the dataset containing the 402 samples (blue), 35 samples (green) and 360 samples (red) as $\mathcal{D},\mathcal{S},\mathcal{W}$ respectively.

\subsection{Machine Learning Input Data Collection}
\begin{figure}
    \centering
    \includegraphics[width=0.9\textwidth]{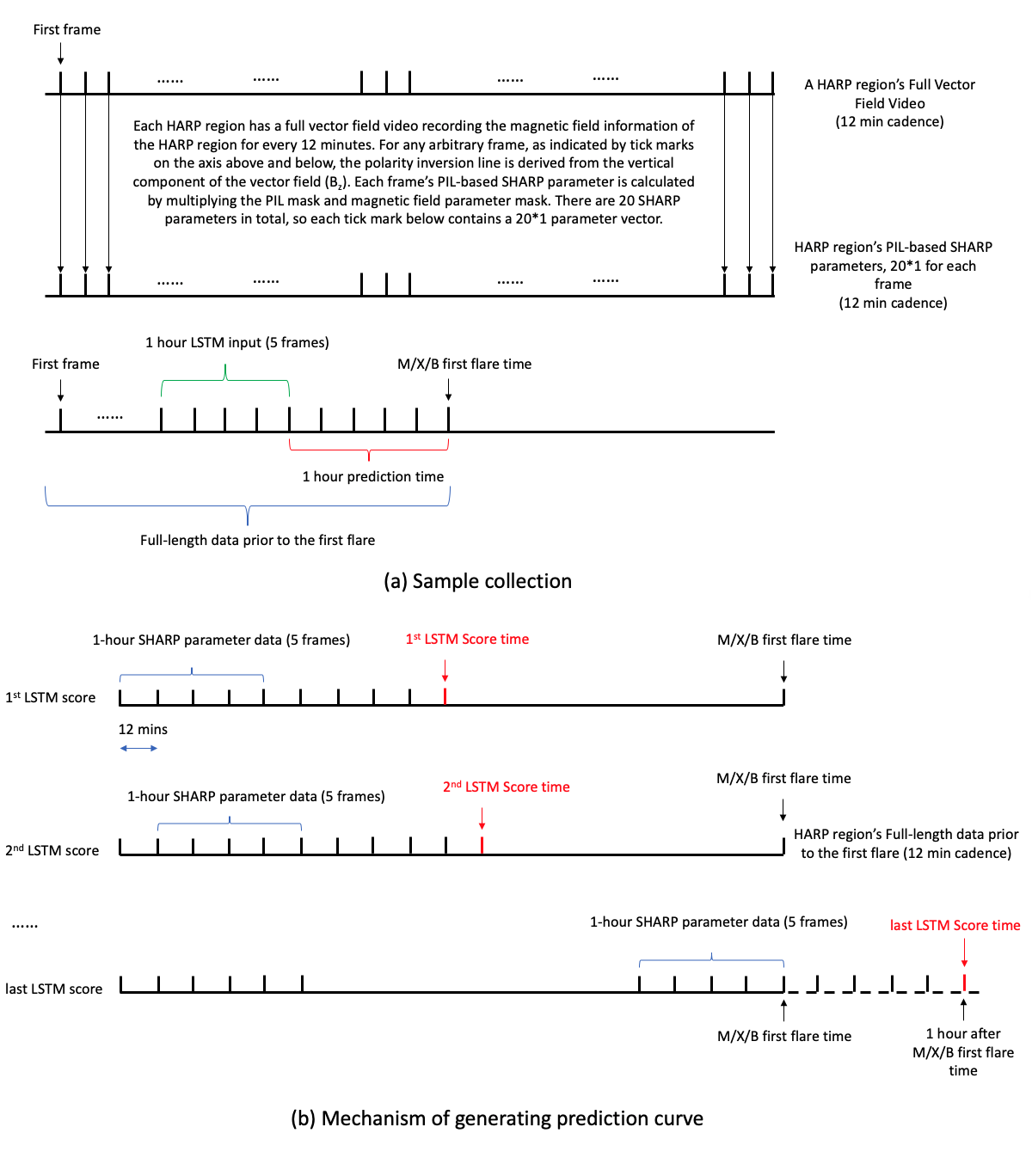}%
    \caption{Schematic showing collection and distillation of data for input to train the LSTM and give the resulting predictions. All data are derived from a time series of three-component HMI vector magnetograms collected from the flare active region patches and saved at a 12 minute cadence. These are down-select to close proximity to the flare original site, the polarity inversion line (PIL). From these data, twenty physical scalar quantities relevant to flare production are calculated, the SHARP parameters.  These data are then grouped together in one-hour, 5-frame sets, which are used to train the LSTM for flare classification. The LSTM model provides the probability of an M/X flare 1 hour after.}%
    \label{fig:collectdata}%
\end{figure}

The machine learning model we employ is the Long-Short-Term-Memory (LSTM) model, which takes multi-dimensional time-series data as the input. In our application, the input data comprise 1-hour time series of the 20 SHARP parameters before each M/X/B first flare, which contains 5 discrete time points saved at a 12 minute cadence.  The time interval for this $20\times 5$ SHARP parameters begins 2 hours before the flare and ends 1 hour before the flare. Consequently, we discard all flare samples which have less than 2 hours of SHARP parameters data preceding them.  For flares with 2 hours of data, we discard all the first flare samples whose preceding vector magnetic field video has more than 10\% missing frames. For all the remaining samples with missing frames, we use linear interpolation to impute the missing values. See (a) of Figure \ref{fig:collectdata} for an illustration on how we select the 1-hour input data.

With all these data pre-processing, we finally obtain a flare list consisting of 97 M/X flares and 305 B flares, with 402 flares (dataset $\mathcal{D}$) in total, which is comparable to the dataset used for first-flare classification in \citep{Yang2019}. Table \ref{tab:flarecount_byyear} provides a flare count summary for each year from December 15, 2010 to June 20, 2018. We label every M or X flare as 1, and every B flare as 0 in the binary classification task. There is only 1 X first flare in 2013 and all other 96 positive class flares are M flares, so we do not further distinguish M and X first flares. We collapse the two classes into a single class. Our deep learning models use the $20\times 5$ SHARP parameters to give a prediction score between 0 and 1, where $0.5$ is the threshold between B and M/X flares.  

\begin{table}[htb]
\centering
\begin{tabular}{|ccccccccccc|}
\hline
\textbf{Class/Year} & 2010 & 2011 & 2012 & 2013 & 2014 & 2015 & 2016 & 2017 & 2018 & Total \\\hline
M/X                   & 0    & 15   & 12    & 20  & 27   & 18    & 3    & 2    & 0    & 97\\\hline
B                  & 2    & 53   & 48  & 52  & 23  & 46  & 52    & 23   & 6    & 305   \\\hline

\end{tabular}
\caption{The number of M/X and B first flares recorded in each year from 2010-2018 in the dataset used for LSTM training and testing.}
\label{tab:flarecount_byyear}
\end{table}

The trained model can then be used to generate a prediction score for a first flare given any 1-hour time series of SHARP parameters. By sliding the 1-hour time window, we can produce a whole path of prediction scores along the entire time series prior to the first flare just as the ones in Figure \ref{fig:4cases}. Figure \ref{fig:collectdata}(b) explains how we use the sliding window approach to collect 1-hour time series of SHARP parameters at a 12 minute cadences to generate a prediction score path given the trained model.

\subsection{Training/Testing Splitting and Normalization}

Dataset $\mathcal{D}$ contains 402 samples coming from 369 different HARP regions, where only 33 HARP regions have both a first M/X flare and a first B flare. In order to give each first flare of a HARP region a complete path of the LSTM classification score, we use the leave-one-out training-test set splitting. Specifically, for each of the 369 HARP regions, we collect all its first flares and associated 1-hour predictor data and put them into the test set while all the other data go in the training set. By doing so, we have 369 different ways of doing training-test set split. 

To normalize each SHARP parameter to have mean zero and variance of one in each of these training-test set pair, we calculate the mean and standard deviation of each SHARP parameter of all samples in the training set only, and then normalize the test set with the mean and standard deviation coming from the training set to avoid any information leaking from the test set to the model training process. With the normalized data, we train 369 LSTM models, each using one of the 369 training-test set pair. Consequently, the classification score path for the first flares of each HARP region is generated with LSTM model trained on data of the first flares from other HARP regions only. 

\section{LSTM results and Sudden Transition}\label{LSTMST}
We apply Long-Short-Term-Memory(LSTM) model \citep{hochreiter1997long} to classify first B flares against first M/X flares, and our model architecture follows that in \cite{Yang2019}. This section details the deep learning neural network architecture and the results of the machine learning model. 

\subsection{Model Description}\label{modeldescription}
\begin{figure}[htb]
\centering
\includegraphics[width=0.9\textwidth]{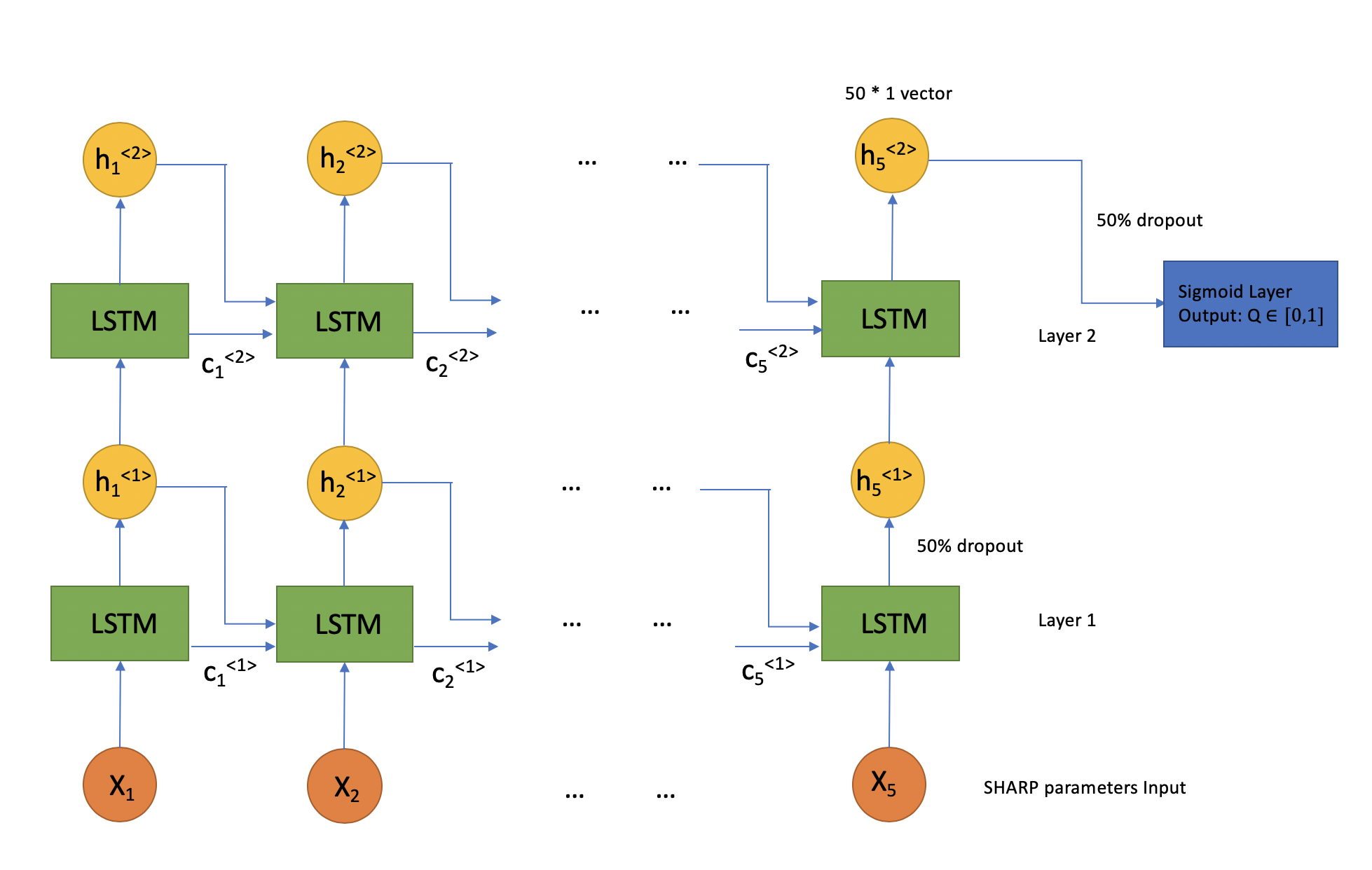}
\caption{Two-layer LSTM architecture. $X_{1},X_{2},\dots,X_{5}$ are the $20\times 1$ normalized SHARP parameters for each frame of the 1-hour input. $h_{1}^{<j>},h_{2}^{<j>},\dots,h_{5}^{<j>}, j=1,2$ are the outputs of each LSTM cell in the first and second layer. $c_{1}^{<j>},c_{2}^{<j>},\dots,c_{5}^{<j>}, j=1,2$ are the memory of each LSTM cell of the first and second layer. All memory and output vectors are $50\times 1$. The output of each LSTM cell of the first layer, after a $50\%$ random dropout, becomes the input of each cell of the second layer. The output vector from the last cell of the second layer, namely the $h_{5}^{<2>}$, is passed to a sigmoid function after $50\%$ random dropout, and gives a prediction score between 0 and 1. The deep learning model is trained with binary cross-entropy loss and Adam optimizer.}
\label{fig:LSTMarch}
\end{figure}

Long-Short-Term-Memory (LSTM) model is a special class of recurrent neural network (RNN), which is widely used to process sequential data such as time-series of stock prices \citep{LSTMstock} and texts in twitters \citep{LSTMtwitter}. The LSTM naturally applies to solar flare prediction in that multi-dimensional time-series data, such as the SHARP parameters, are the typical predictors. An RNN uses a recursive, highly non-linear mathematical operation to process the sequential inputs. The LSTM improves some of the drawbacks of traditional RNN by introducing a separate memory (called the cell state) to the neural network, where data of each time point in the sequential input can decide how much information the previous time points provided to the neural network can be kept for future use. The cell state enables the information of data coming from the early stage to have an effect even after a very long time. Figure \ref{fig:LSTMarch} shows the tensor flows in our deep learning model consisting of LSTM cells. The input of our neural network is the normalized 1-hour SHARP parameter time-series and the output is a classification score between 0 and 1.

\subsection{LSTM Results: Classification Score Paths}

\begin{figure}[htb]
    \centering
    \includegraphics[width = 0.9\textwidth]{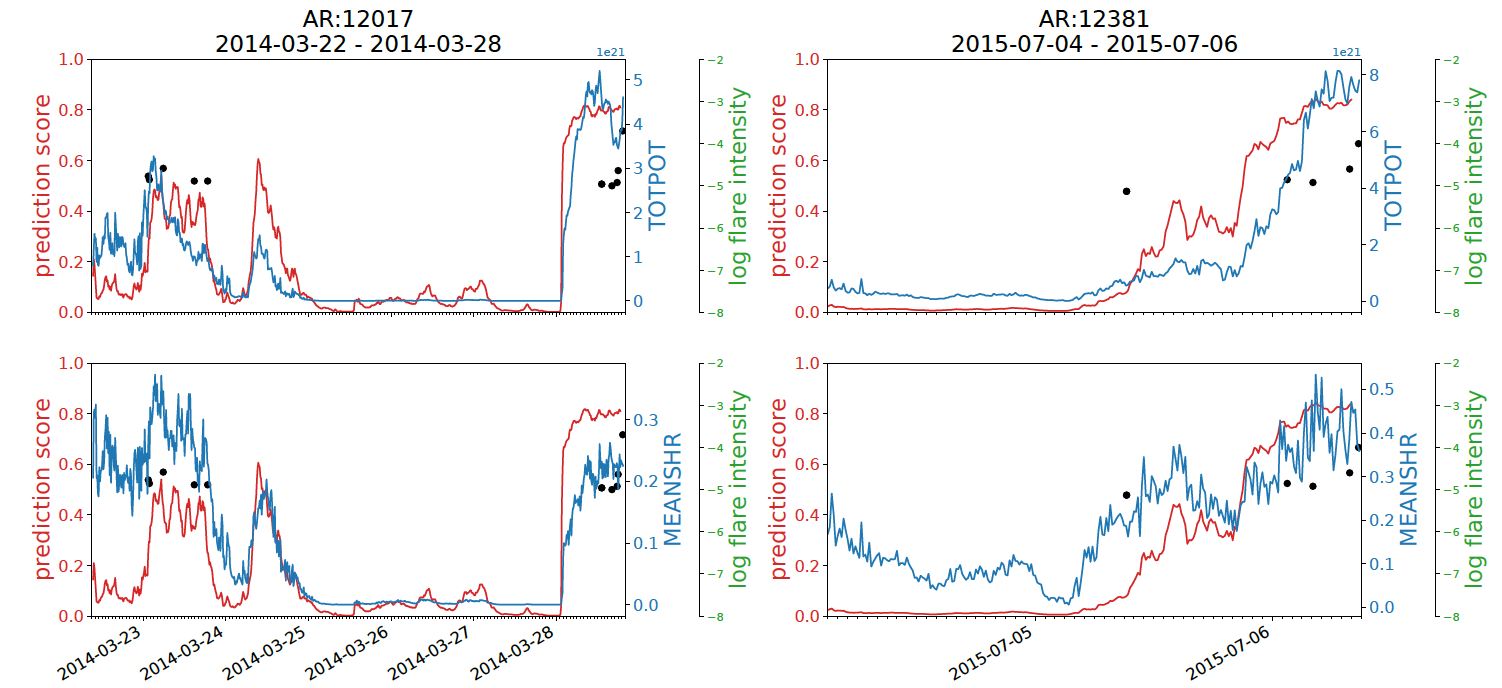}
    \caption{Leave-one-out prediction score and time-series of key SHARP parameters of the first M flare of AR 12017 and 12381. Blue curves show the time-series of TOTPOT and MEANSHR, while red curves show the leave-one-out prediction score. Black dot show recorded flare events from the given active regions with the height proportional to the common logarithm of flare intensity found on the ``$\log$ flare intensity" axis. Minor ticks on the time axis are plotted for every hour. The date format here and all similar figures is Year-Month-Day. Units for total potential energy (TOTPOT) and mean shear angle (MEANSHR)  are erg $\cdot \mathrm{cm}^{-1}$ and degree respectively. Both SHARP parameters are weighted by the PIL mask. 
    }
    \label{fig:exampleofar}
\end{figure}

For each of the 369 pairs of training set and test set, we produce a trained LSTM model as shown in Figure \ref{fig:LSTMarch}. The LSTM gives us a complete prediction score path for every first flare in the test set that we call the \textbf{leave-one-out prediction score} for the first flare. Figure \ref{fig:exampleofar} shows the leave-one-out prediction score and two SHARP parameters for the first M flare of AR 12017 and 12381. The unit of total free energy density (TOTPOT) is erg $\cdot\mathrm{cm}^{-1}$, and the unit of mean shear angle (MEANSHR) is degree.  Each black dot represents a flare event with the logarithm-transformed intensity shown on the axis ``$\log$ flare intensity" (green color). The same intensity scale applies to all similar graphs in the following sections. For the sake of formatting simplicity, we do not draw the ``$\log$ flare intensity" axis in the subsequent similar graphs. It also can be seen that the SHARP parameters and prediction scores are higher when a stronger flare is upcoming.

We see that the red curve (leave-one-out prediction score) and the blue curve (SHARP parameters) in Figure \ref{fig:exampleofar} have some non-negligible positive correlations. The case for AR 12017 represents the type of prediction score path with sudden transitions. It can be seen that starting from the midnight of March 28, 2014, the prediction score suddenly increased to a very high level in a few hours, together with the sudden increase of both features. It can be easily inferred that the polarity inversion line emerged in a rather short period for AR 12017 during this time. During March 26$^{\mathrm{th}}$ to 28$^{\mathrm{th}}$, both SHARP parameters remain constantly at zero because the vector field during this period has no polarity inversion line detected. For AR 12381, the prediction score transitioned from relatively low to fairly high at a much slower pace. However, it is common in both active regions that there are times of very low LSTM prediction score followed by periods of high LSTM prediction score. 

One may also notice that the MEANSHR is more locally volatile than TOTPOT despite the fact that both of them positively correlate with the prediction score path. This is because the total energy density is calculated as the average of the energy of all pixels, weighted by the PIL mask. The mean shear angle is the sum of shear angle of all pixels, weighted by the PIL mask, and then divided by the number of all non-zero pixels in the PIL mask. Therefore, the MEANSHR is not weighted by field strength. 

The high prediction score and low prediction score of a single HARP region acts as a "contrasting pair", making it possible for one to analyze the physical mechanism of flare eruption learnt by the LSTM model within an active region. Instead of comparing the SHARP parameters of AR 12017 during its high LSTM score time against the SHARP parameters of AR 12381 during its low LSTM score time, it makes more intuitive sense for us to compare the SHARP parameters within an active region. So among all HARP regions we are particularly interested in those that have both a low and a high LSTM score period like AR 12017 and 12381 and all regions in Figure \ref{fig:4cases} where within-AR paired comparison is possible. 
 
To give such a prediction score pattern a more formal definition, we define that an active region's prediction score path has gone through a \textbf{sudden transition} if:
\begin{itemize}
    \item An M/X flare happened at the end of the leave-one-out prediction score path.
    \item There is a certain time when the LSTM prediction score is above 0.7, and persist for at least 36 minutes afterwards. We call this time the \textbf{post transition time}.
    \item Prior to the post transition time, if there is any time when the LSTM prediction score is below 0.3, and persist for at least 36 minutes afterwards. We call this time the \textbf{prior transition time}.
\end{itemize}

Table \ref{tab:ST} in the appendix provides all 35 cases where we found a sudden transition along their leave-one-out prediction score paths. Since the prediction score pattern for these active regions is quite similar which might indicate that the underlying physical processes may also be similar, the cases in the table can be used in other researches when one wants to give a uniform explanation for strong first flare eruption.

In all the cases in Table \ref{tab:ST}, the leave-one-out LSTM prediction score path has a certain time range where the prediction score jumped from a low level to a high level. Some of the transitions happened quickly, such as in AR 12017 (2.5 hours) and AR 12182 (4 hours), but slower in cases such as in AR 11718 (50 hours). In the next section, we will explain our adopted statistical technique to capture the driving force of sudden transitions of LSTM scores, and we will present case studies to illustrate how certain SHARP parameters have changed significantly before and after the LSTM score transition.

\section{Interpretation of the LSTM Results}\label{LSTMINT}

To obtain interpretations of the LSTM predictions, we perform a clustering analysis designed to extract the contrasts in SHARP parameters that ultimately distinguish weak flares from strong flares. In our analysis, two clusters are defined by high and low LSTM prediction score paths that corresponding to weak and strong flares classes respectively. We perform a clustering analysis on all $20\times 5$ SHARP parameter time series that generated the prediction scores. To envision our application of cluster analysis, consider an example where one input of $20\times 5$ SHARP parameters gives an LSTM score at 0.05 and and a second $20\times 5$ set of time series data gives a score of 0.95. If both sets of time-series are very similar except for a single SHARP parameter A, one would conclude that SHARP parameter A is the feature  determining the LSTM prediction score differences. More generally, we aim at finding a subset of all 20 SHARP parameters that can define an appropriate similarity measure between the time-series inputs. The chosen similarity measure will be considered appropriate if inputs that generate low/high LSTM scores are more similar to each other than to the inputs that generate high/low LSTM scores.

Our clustering analysis requires a measure of similarity which we develop in three essential steps. First, we use a cubic spline to fit all SHARP parameters and derive their first-order time derivatives. Second, we apply the dynamic time warping (DTW) distance to all SHARP parameters and their derivatives to construct a coarse similarity measure. DTW is chosen because our input data is of the form of multi-dimensional time series data with strong temporal dependency. Third, we finalize the construction with principal component analysis, show that the constructed measure is appropriate with visualizations and then derive our interpretations. In Section~\ref{subsec:dtw_construct}, we describe how we construct DTW features that differentiate high LSTM prediction score paths from the low LSTM prediction score paths, i.e. the strong flares from weak flares.

\subsection{A Coarse Similarity Measure based on Dynamic Time Warping and Spline Fitting}
\label{subsec:dtw_construct}

Dynamic Time Warping (DTW) distance \citep{DTW_paper} is a measure of similarity between any two sets of 1-d sequential data, especially temporal data. It is widely used in situations such as speech recognition where speakers might have different talking speed in various recordings \citep{DTWspeech}, and gene expression time-series analysis where different biological processes might unfold with different rates \citep{DTWgene}. In general, DTW measures how similar two time-series are to each other: the smaller the DTW distance is, the more similar the two time-series are. We use DTW distance to measure the distance between time-series inputs into the LSTM for two reasons: First, the DTW distance is a very powerful yet simple distance measure of time-series data that has been in use for the past few decades, especially in the task of time-series classification \citep{xi2006fast,xing2010brief,wang2013experimental}. Second, the DTW distance can capture similar patterns of evolution occurring at different time scales that lead to strong flares.

\begin{figure}[htb]
    \centering
    \includegraphics[width = 0.9\textwidth]{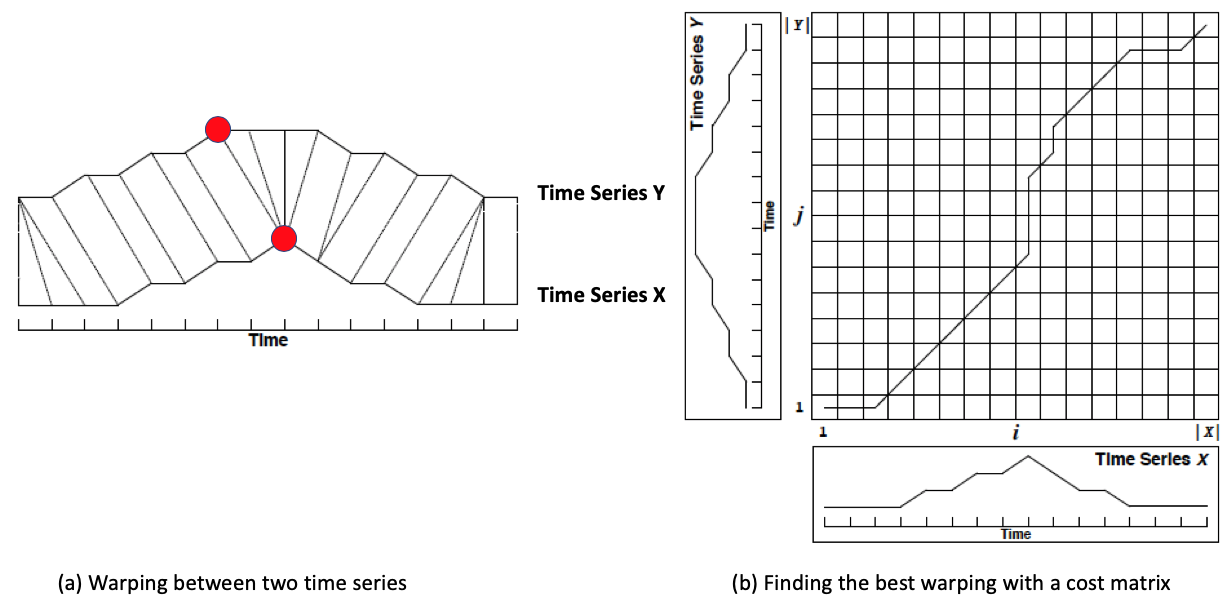}
    \caption{Explanation of Dynamic Time Warping. (a) shows the pairing of time points under the dynamic time warping distance measure. (b) shows that dynamic time warping is essentially a dynamic programming problem inside the pair-wise distance matrix of two time-series. Figures are adopted from \cite{salvador2004fastdtw}.}%
    \label{fig:dtw}%
\end{figure}

Figure \ref{fig:dtw}(a) shows the motivation of dynamic time warping. There are two time-series, X and Y with the same time domain. One can see that the increasing trend of X and Y are asynchronous, where X takes a bit longer to reach its peak. The Euclidean distance between X and Y calculates the differences of each pair of points at each time point. But it fails to account for the similarity of the time evolution of X and Y. Dynamic time warping optimizes the pairing of time points so that the two increasing trends culminating at the two red points can be paired with each other. Figure \ref{fig:dtw}(b) illustrates how to find the optimal pairing. With the two time-series X and Y, one could simply calculate a cost matrix where each entry $(i,j)$ contains the cost of pairing $X_i$ and $Y_j$. Typically, the cost of pairing $X_i$ and $Y_j$ is $(X_{i}-Y_{j})^2$. With the constraint that the first and last point of X should be paired with the first and last point of Y, finding the optimal pairing between X and Y is equivalent to finding a path from the lower left corner to the upper right corner in the cost matrix with the least cost.

\begin{figure}[htb]
    \centering
    \includegraphics[width = 0.9\textwidth]{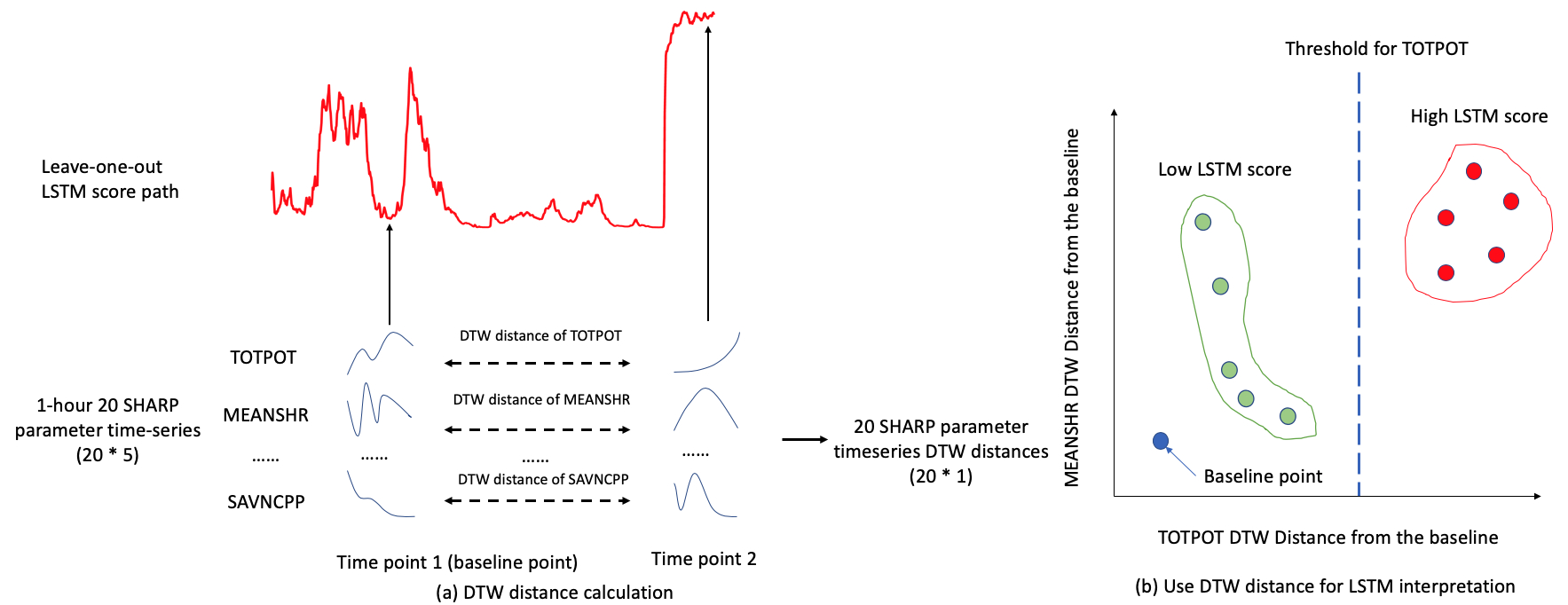}
    \caption{Illustration of how we use dynamic time warping distance to compare any two LSTM inputs and do interpretation. (a) For any two arbitrary time $t_1$ and $t_2$ on the prediction score path of a certain HARP region, we collect their corresponding 1-hour PIL-based SHARP parameter time-series $\textbf{X}_{t_1}$ and $\textbf{X}_{t_2}$ and calculate the DTW distance between the 1-hour time-series of each of the 20 SHARP parameters of $\textbf{X}_{t_1}$ and $\textbf{X}_{t_2}$. Illustrations of the time profiles of parameters TOTPOT, MEANSHR and SAVNCPP are provided. The final similarity between $\textbf{X}_{t_1}$ and $\textbf{X}_{t_2}$ is a $20\times 1$ vector $\textbf{d}(\textbf{X}_{t_1},\textbf{X}_{t_2})$. In practice, we define a time point $t_0$ as the baseline point for the score path where the prediction score is the lowest along the path: $t_0=\argmin_t \mathrm{score}(t)$. We then calculate the corresponding DTW distance between the LSTM inputs at any time point $t$ and the baseline point $t_0$. (b) We show an ideal case where interpretations could be derived. The baseline point is the blue point, and the DTW distances of SHARP parameter TOTPOT and MEANSHR of other time points against the baseline are plotted in a 2-d space. We aim to find the important variables of LSTM prediction by finding which variables can threshold low and high score inputs with the pattern shown in (b).}
    \label{fig:DTW_Explained}
\end{figure}

Figure 7(a) shows how DTW distance calculation is applied to any pair of $20\times 5$ LSTM inputs at two arbitrary time points of a score path. For any two $20\times 5$ LSTM inputs, we calculate the DTW distance between the time-series of each of the 20 SHARP parameters. This leads to a $20\times 1$ vector where each dimension is the time-series similarity of a given SHARP parameter. 
Rather than compare all possible pairs of LSTM inputs for 402 flare samples in data set $\mathcal{D}$, we fix a \textbf{baseline point} for every leave-one-out prediction score path, and compare the LSTM inputs of any prediction score path against the LSTM inputs at the baseline point of the path. This step significantly reduces the amount of pairs of LSTM inputs for DTW distance calculation. We choose the baseline as the time point where the prediction score is the lowest. Panel (a) of Figure \ref{fig:DTW_Explained} shows an example where time point 1 is the baseline point of the score path. Figure \ref{fig:DTW_Explained}(b) shows how this DTW distance can be potentially helpful for interpreting LSTM prediction. Given the baseline point (shown in blue), we plotted several other points in a 2-d space where the coordinates are DTW distance of their 1-hour TOTPOT and MEANSHR from the 1-hour TOTPOT and MEANSHR of the baseline. If we can find a few SHARP parameters such as the TOTPOT in the hypothetical case in Figure \ref{fig:DTW_Explained}(b), where DTW distance can linearly separate low and high LSTM score cases, then we can establish that when the time evolution of these SHARP parameters becomes sufficiently different from the baseline (crossing the decision boundary), we can anticipate the LSTM score will be high, thus indicating a strong flare. 

\begin{figure}[htb]
    \centering
    \includegraphics[width=0.9\textwidth]{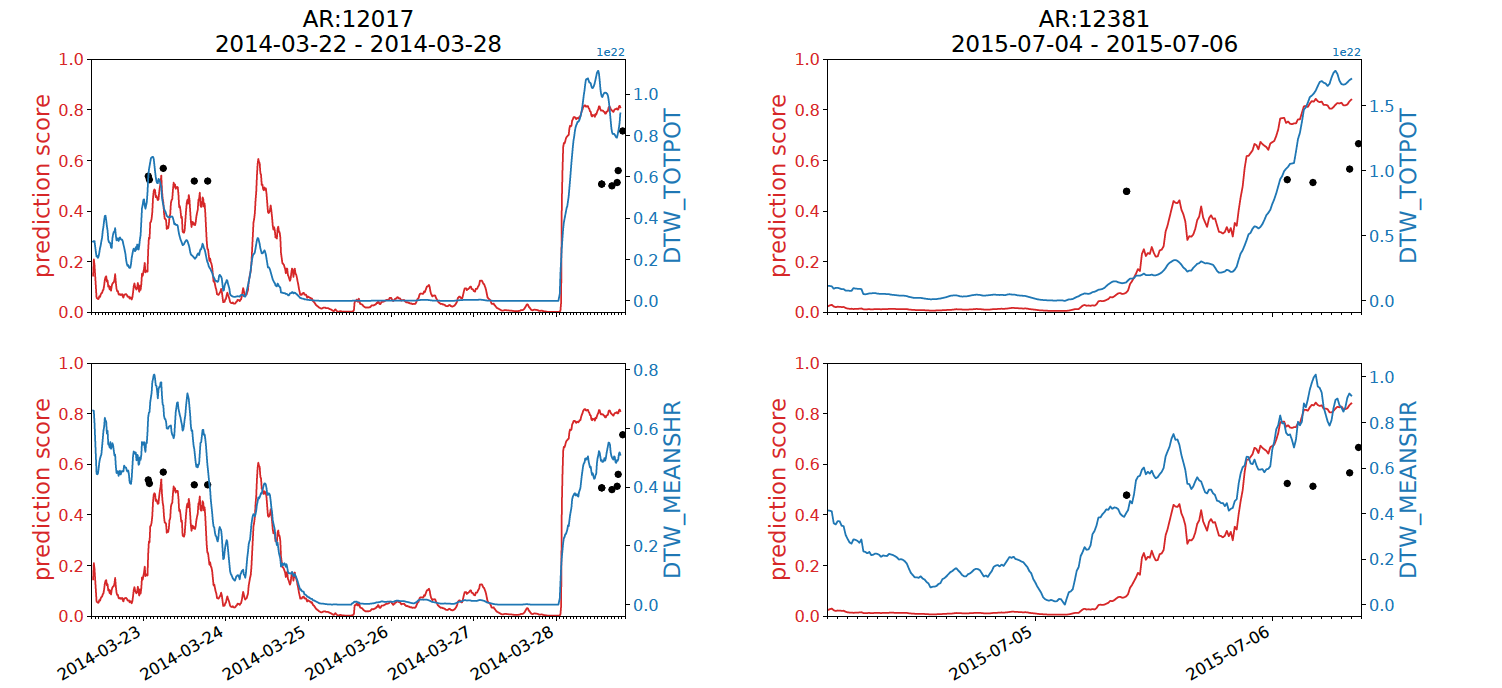}
    \caption{Prediction scores and DTW distance time-series of TOTPOT and MEANSHR for AR 12017 and 12381. Blue curve shows the time-series of DTW distance of SHARP parameter TOTPOT and MEANSHR prior to the first M/X flare, red curve shows the leave-one-out prediction score. All formats in the figure follow that in Figure \ref{fig:exampleofar}. Note the strong correspondence between the DTW distances and the prediction scores.}
    \label{fig:DTW_example}
\end{figure}

The 20-dimensional DTW distance measure for the leave-one-out score path is a useful multi-dimensional similarity measure because as we will show, it is highly correlated with the LSTM prediction scores. In Figure \ref{fig:DTW_example}, we show the constructed DTW distance time-series of the total free energy (per unit length in the radial direction) (TOTPOT) and the mean shear angle (MEANSHR) for AR 12017 and 12381, along with their prediction score paths. Comparing the time series of TOTPOT and MEANSHR in Figure \ref{fig:exampleofar} with their DTW distance features Figure \ref{fig:DTW_example}, one can see that the DTW distance features are smoother and have high correlation with the strong-flare prediction score curve.

While the DTW is useful for our analysis, there are limitations that must be addressed.  First, while the DTW distance has the ability to capture the asynchronous trends, but this has the drawback that it may modify the time derivatives of time-series to yield a better corresponding match. However, the growth rate information of SHARP parameters are physically significant and potentially vital to the LSTM. To compensate for this limitation, we directly calculate the time derivatives of the SHARP parameters. Second, we need to address the local volatility of SHARP parameters. As one can see from Figure \ref{fig:exampleofar}, the MEANSHR of AR 12017 and 12381 are highly volatile during high LSTM score times, but the TOTPOT of AR 12381 is not as volatile. So we must also compare the volatility of two LSTM inputs alongside their general trend and slopes.

\begin{figure}[htb]
    \centering
    \includegraphics[width=0.9\textwidth]{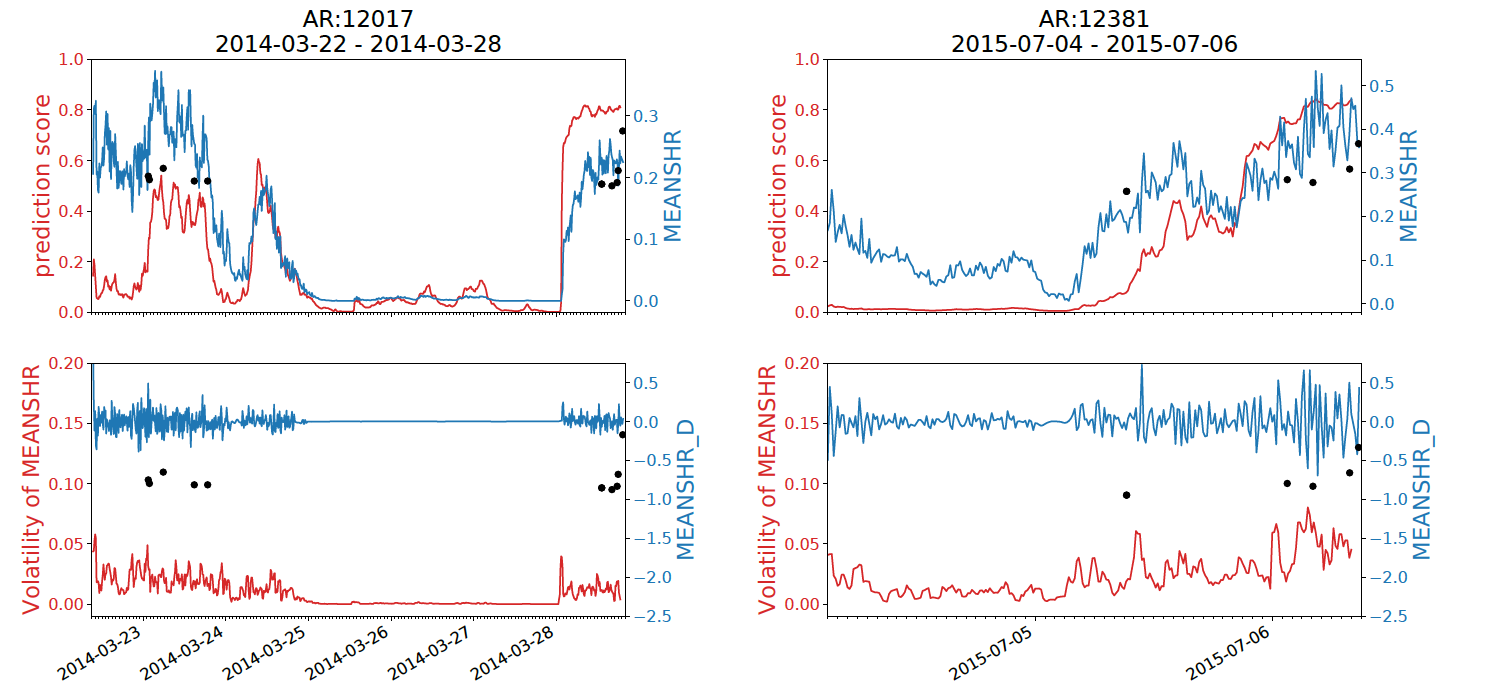}
    \caption{Plots of MEANSHR time-series and the first-order derivatives for AR 12017 and 12381. Blue curves show the time series of MEANSHR and its first-order derivative prior to the first M/X flare in the top and bottom panels respectively. The first-order derivative is obtained by fitting a natural cubic spline on the MEANSHR time-series. Red curves show the leave-one-out prediction score and the 1-hour local standard deviation of MEANSHR in the top and bottom panels respectively. All formats of the figure follow that in Figure \ref{fig:exampleofar}. Unit of MEANSHR\_D is degree $\cdot\mathrm{h}^{-1}$.}
    \label{fig:exampleofar_derivative}%
\end{figure}

To better account for slope and local fluctuations, we use the natural cubic spline \citep[Chapter~5]{ESL} to interpolate the SHARP time series data from which we derive their first-order derivative time-series. Examples for of AR 12017 and 12381. are shown in Figure \ref{fig:exampleofar_derivative}. Here, the original (non-interpolated) MEANSHR, the derivative of MEANSHR (MEANSHR\_D) and the leave-one-out prediction score path, while the bottom panels show the moving 1-hour standard deviation of MEANSHR (1-hour local volatility of MEANSHR). It can be seen that derivative based on our spline fitting of the MEANSHR is capturing more about local volatility than the large-scale slope of MEANSHR.  When the MEANSHR is increasing, we can see some significant non-zero derivatives in MEANSHR\_D. But instead of giving a consistently positive derivative, the MEANSHR\_D is fluctuating wildly around zero, with both strongly positive and strongly negative values. Also, the magnitude of fluctuations of the MEANSHR derivatives is strongly correlated with the local volatility of the MEANSHR time-series expressed in the standard deviation. Most importantly, the MEANSHR fluctuates strongly when a solar flare is happening or is about to happen, and MEANSHR\_D is very volatile prior to flares with high intensity.

In general, the volatility of the MEANSHR and similar SHARP parameters comes from a combination of noise, reconstruction ambiguity and inherent variability. Over a period of 12 minutes, the magnetic vector of the weak field pixels may change wildly. For the PIL-based SHARP parameters, there is an additional source of local volatility, which is the unstable PILs found in adjacent frames. The algorithm in \cite{Jingjing2019} does not account for the time-consistency of PIL across frames but detect PIL frame by frame. As a result, the PIL mask of the current time can be very different from the PIL mask 12 minutes later. The unstable PIL introduces additional volatility in the calculation of all SHARP parameters, especially the gradient-related parameters. The energy-related parameters are less affected because the free energy of each pixel is generally more stable and locally continuous than the gradient of each pixel.

\begin{figure}[htb]
    \centering
    \includegraphics[width = 0.9\textwidth]{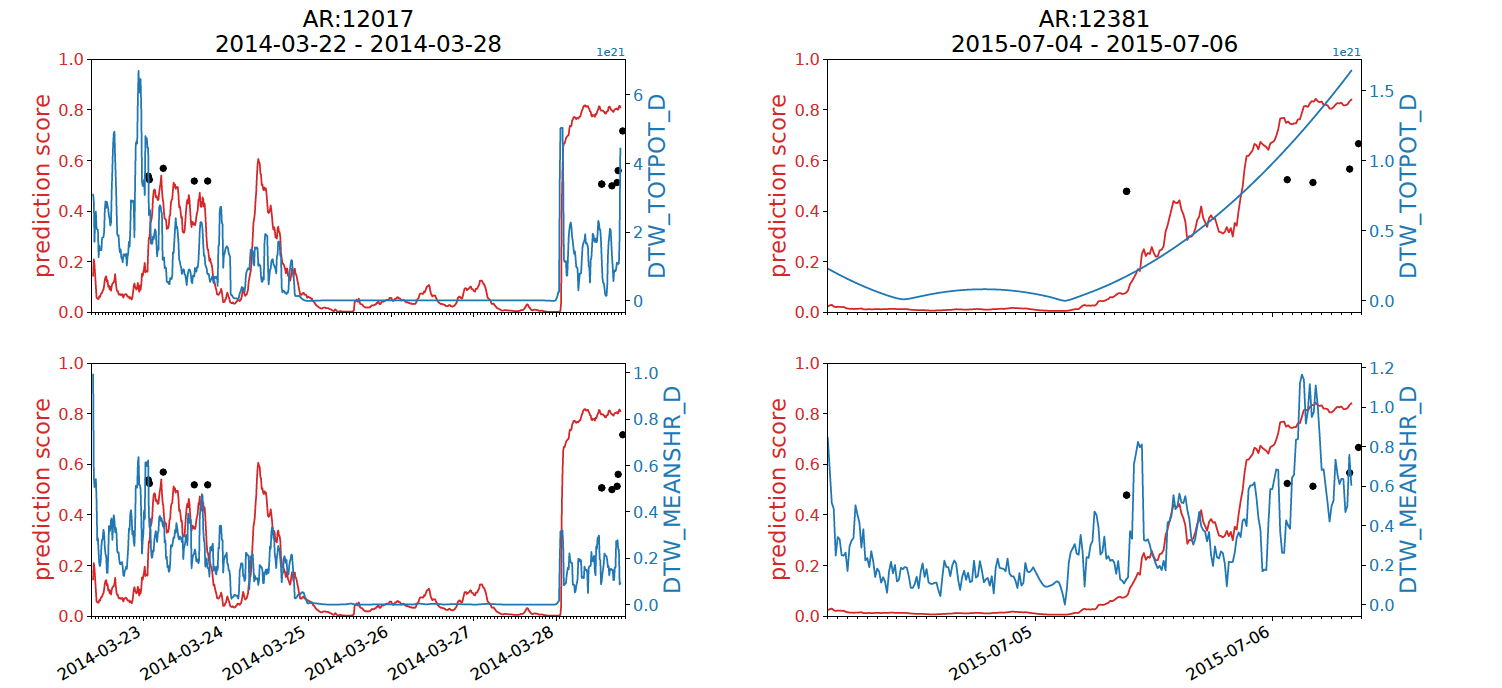}
    \caption{Plots of DTW distance for the first-order derivatives of TOTPOT and MEANSHR for AR 12017 and 12381. Blue curve shows the time-series of DTW distance of the derivatives of SHARP parameters TOTPOT and MEANSHR prior to the first M/X flare, red curve shows the leave-one-out prediction score. All formats of the figure follow that in Figure \ref{fig:exampleofar}. Unit of DTW\_TOTPOT\_D is erg $\cdot\mathrm{cm}^{-1}$ $ \cdot\mathrm{h}^{-1}$. Unit of DTW\_MEANSHR\_D is degree $\cdot\mathrm{h}^{-1}$. We find a very close similarity between the evolution of the prediction scores and the DTW distance of the time derivatives of TOTPOT and MEANSHR.}
    \label{fig:DTWDeri_example}%
\end{figure}

We now calculate the DTW distance measure for time derivatives in the exact same fashion as used for the SHARP time-series, including spline fitting, sliding window method (see Figure \ref{fig:collectdata}(b)), and comparison to a chosen baseline point. In Figure \ref{fig:DTWDeri_example}, we show the DTW distance feature of the time derivatives of the total free energy density (TOTPOT) and mean shear angle (MEANSHR). We again find a very close correspondence between the prediction score and the DTW distance. The DTW measure of the derivative of TOTPOT for AR 12381 is in particular smooth compared to the other three examples and rises in unison with the prediction score. This example again shows that the energy-related SHARP parameters are generally less volatile than gradient-related ones. 

The DTW measure is calculated for SHARP parameters and their time derivatives (except for X\_SIZE, Y\_SIZE), which provides a $36\times 1$ dimensional similarity measure calculated at all time points along the leave-one-out score paths of the 402 flare samples of dataset $\mathcal{D}$. We discard all samples whose baseline point has an LSTM score above 0.2 for they may have already contain some signals of flare eruption.  We choose the cutoff point at 0.2 instead of 0.3 used to define sudden transition because typically when LSTM score is close to 0.3, it starts to increase rapidly and is no longer considered as a ``quiet'' time of the region. After this thresholding, we obtained dataset $\mathcal{W}$ with appropriate baselines, containing 70 M/X flares and 290 B flares. All samples in dataset $\mathcal{S}$ are in dataset $\mathcal{W}$. 

 The DTW measure is far from ideal since it is both multi-dimensional and non-orthogonal with some shared information of flare eruptions. In the next subsection, we will finalize our construction of the similarity measure by reducing the dimension using principal component analysis. We will also illustrate the constructed measure and derive interpretations of LSTM model. 

\subsection{A PCA analysis of DTW distance features}\label{PCAanalysis}

In this subsection, we apply Principal component analysis (PCA) to the DTW distance features in order to highlight principal components that can distinguish a high probability of large flares. PCA is a dimension reduction technique that uses orthogonal transformations to convert a set of observations of possibly correlated variables into a set of linearly uncorrelated variables called principal components. The orthogonal transformation is defined in such a way that the first principal component has the largest possible variance, and each succeeding component in turn has the highest variance possible under the constraint that it is orthogonal to the preceding components. More details on PCA can be found in \cite[Chapter~12]{Bishop2006}.

\begin{figure}[htb]
\centering
\includegraphics[width = 0.9\textwidth]{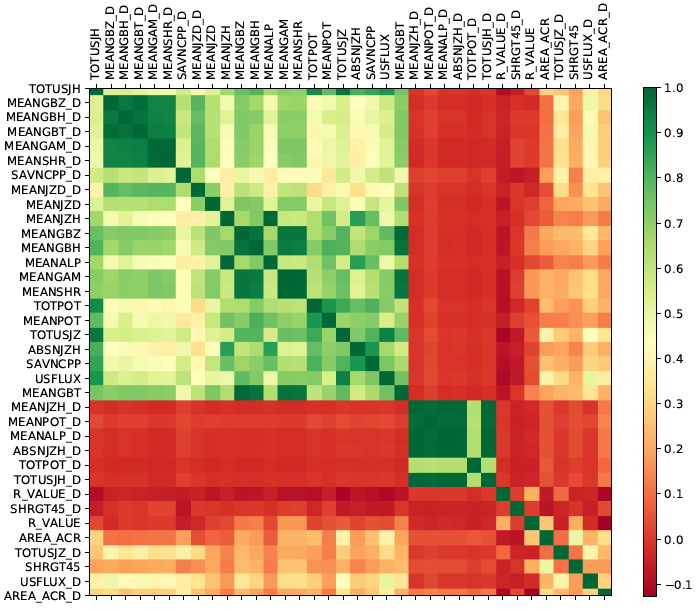}
\caption{Pearson correlations among 36 DTW distance features of samples of dataset $\mathcal{W}$. There are two highly correlated blocks in dark green, one consisting of the DTW distance feature of time derivatives of MEANGBZ, MEANGBH, MEANGBT, MEANSHR, MEANGAM, and the other including the DTW distance feature of the time derivatives of MEANJZH, MEANPOT, MEANALP, ABSNJZH, TOTUSJH. Furthermore, there is a weakly correlated block that includes more than 20 features on the top left corner of the correlation matrix (from TOTUSJH to MEANGBT). More quantitative description about the correlation structure can be found in Figure \ref{fig:PCAresults}.}
\label{fig:DTWcorr}
\end{figure}

In our analysis, we next perform PCA eigenvalue decomposition of the Pearson correlation matrix of 36 DTW distance features across all samples in dataset $\mathcal{W}$, which are shown in Figure \ref{fig:DTWcorr}. This correlation matrix shows a large group of DTW distance features that are correlated with each other shown in green. A highly correlated group is found in the upper left corner of the matrix consisting of  MEANGBZ\_D, MEANGBH\_D, MEANGBT\_D, MEANGAM\_D and MEANSHR\_D. A high correlation between SHARP parameters or their derivatives means that their time evolution away from the baseline value will occur in a similar way while being very different from the time evolution of other distinct groups of highly correlated parameters. Recall that the LSTM score path moves in the same direction with the DTW distance features of TOTPOT, MEANSHR in Figure \ref{fig:DTW_example}, suggesting that these two quantities have some similar influence on the LSTM prediction score. The overlapping information between these two dimensions for distinguishing weak from strong flare potential is expressed with a Pearson correlation value of approximately 0.7.

Indeed, if any DTW distance feature becomes significantly different from the baseline as the LSTM score becomes very high, we can expect all highly correlated features will be very different from the baseline as well. As a result of the correlation structure, we can analyze blocks of highly correlated DTW distance features instead of individual DTW distance features. This idea
leads us to apply principal component analysis (PCA) to do dimension reduction to account for the shared information among the DTW distance features.

\begin{figure}[htb]
\centering
    \includegraphics[width = 0.9\textwidth]{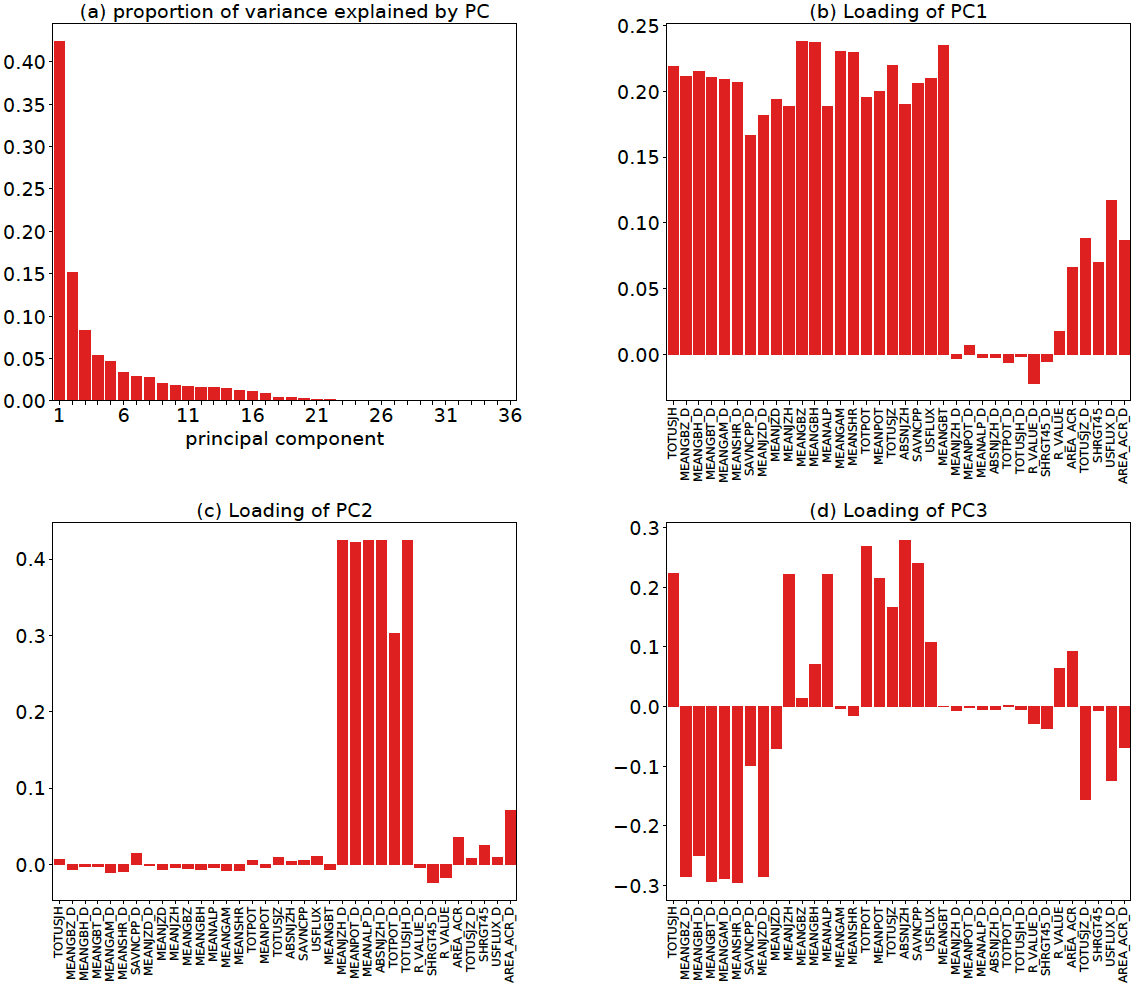}
    \caption{Results of Principal Component Analysis (PCA) on 36 DTW distance features of samples from dataset $\mathcal{W}$. The order of the variable on x-axis of (b), (c) and (d) is the same as the rows in Figure \ref{fig:DTWcorr}. (a) shows the proportion of variations explained by each of the 36 principal components of the PCA. (b), (c) and (d) show the variable loadings of the first, second and third PCs with the largest eigenvalues. PC1 captures the large green block with over 20 features in upper left corner of Figure \ref{fig:DTWcorr}, PC2 captures the small green block in the lower right corner and PC3 is the difference between variables highly correlated with TOTPOT and variables highly correlated with the time derivatives of MEANSHR.}
\label{fig:PCAresults}
\end{figure}

Figure \ref{fig:PCAresults} shows the main results of PCA on the 36 DTW distance features. Panel (a) shows the proportion of variations explained by each of the 36 principal components, and (b), (c), (d) shows the feature loadings of the top 3 principal components. The first principal component (PC1) has large and relatively uniform loadings on all variables in the large green block on the upper left corner of Figure \ref{fig:DTWcorr}. PC1 is a combination of all correlated DTW distance features of the large green block. When PC1 becomes large for a certain HARP region at some time, we can expect many of the features with large principal component loadings in PC1, such as TOTPOT and MEANSHR\_D, have become significantly different from the baseline. PC2 captures the smaller dark green block of variables in the lower right corner of Figure \ref{fig:DTWcorr}, including the derivatives of several SHARP parameters. PC3 is a combination of the DTW features of many SHARP parameters (TOTUSJH, MEANJZH, MEANALP, TOTPOT, ABSNJZH, SAVNCPP) subtracting the DTW features of the derivatives of a subset of other SHARP parameters (MEANGBZ\_D, MEANGBH\_D, MEANGBT\_D, MEANGAM\_D, MEANJZD\_D and MEANSHR\_D).

To interpret the PCA results, one can take PC1 as the component containing the information of the changes of the trend of the time evolution of many SHARP parameters. PC2, as shown in Figure \ref{fig:DTWDeri_example}, contains the slope information of SHARP parameters such as TOTPOT, TOTUSJH and ABSNJZH, which are typically the less noisy SHARP parameters. And PC3, which turns out to be the most interesting principal component based on the post hoc analysis in the following subsections, is the difference between the time evolution of the trend of parameters such as TOTPOT, ABSNJZH, SAVNCPP and USFLUX and the local volatility of MEANSHR, MEANGBZ, MEANGBH, MEANGAM, and MEANGBT. To put it simply, PC3 is to subtract noises of some SHARP parameters from signals of other SHARP parameters. To fully evaluate the significance of the variable loadings in the principal components, we will show how SHARP parameters actually change during sudden transitions using case studies and classification tree model.

\begin{figure}[htb!]
    \centering
    \includegraphics[width = 0.7\textwidth]{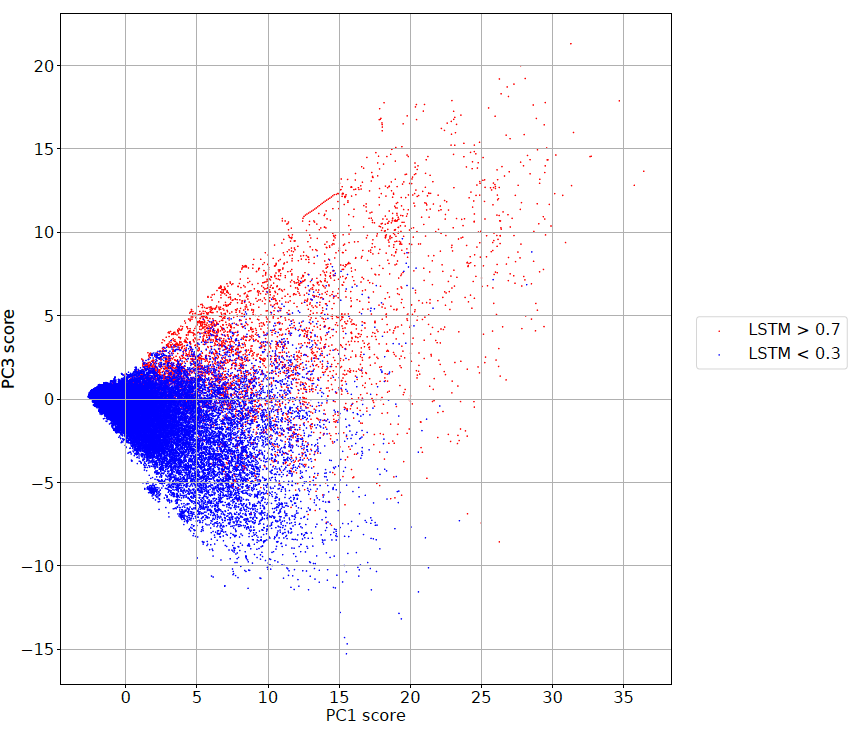}
    \caption{PC1 and PC3 score of cases where LSTM score is greater than 0.7 (red) and lower than 0.3 (blue). We found that typically, inputs with LSTM score higher than 0.7 will have positive PC3 score, while inputs with LSTM score lower than 0.3 will have negative PC3 score.}
    \label{fig:pc13_score}
\end{figure}

We next calculate the principal component scores for the 36 DTW distance features for all samples from the dataset $\mathcal{W}$ to determine whether they can distinguish low and high LSTM score cases. In Figure \ref{fig:pc13_score}, we plot all LSTM inputs resulting in a prediction score below 0.3 (blue) or above 0.7 (red), in a 2d space with the x-axis being their PC1 score and the y-axis being their PC3 score. It can be seen in Figure \ref{fig:pc13_score} that high LSTM score cases are typically those whose PC3 score is positive, and most are larger than 3. On the contrary, the low LSTM score cases are those with low PC3 scores. In terms of the PC1 score, even though there are many cases with high PC1 score from both classes, there are many cases with high LSTM score that have very large PC1 score, well beyond 20, which is rarely seen in low LSTM score cases.

\begin{figure}[htb]
\centering
\includegraphics[width = 0.9\textwidth]{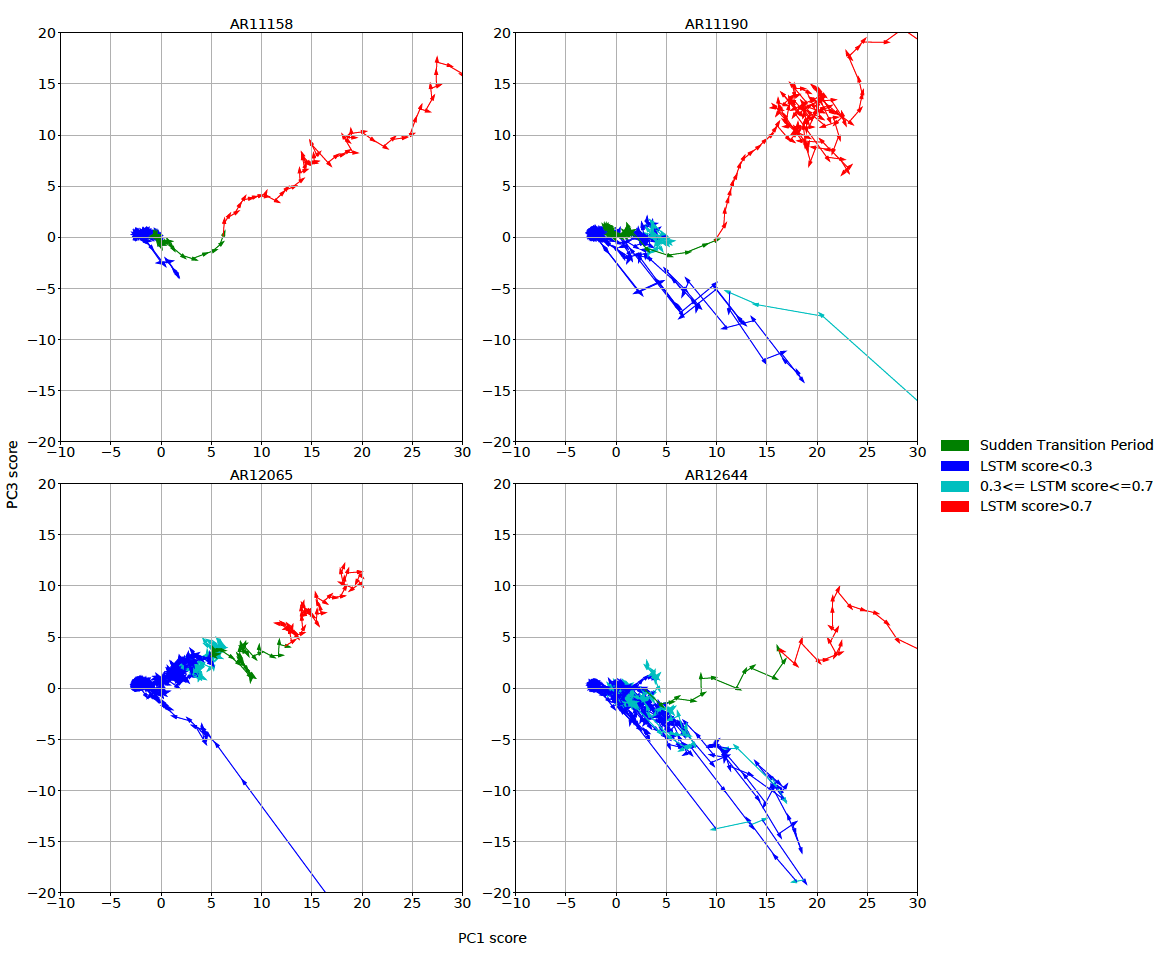}
\caption{Four case studies on AR 11158, 11190, 12065 and 12644. Evolution path of PC1 and PC3 are plotted with connected arrows showing the direction of changes for every 12 minutes. Green arrows represent sudden transition period, where LSTM scores changed from very low ($<$ $0.3$) to very high ($>$ $0.7$). Other arrows are during the non-transition times, LSTM score below 0.3, between 0.3 and 0.7 and above 0.7 are colored blue, cyan and red, respectively. We find that: (i) during the sudden transition time, the PC1 and PC3 scores of each active region become further away from the origin and travel towards the places where both PC scores are high. (ii) It seems that the location of the sudden transition period in the 2-d PC score space is similar across active regions, perhaps there exists a threshold in the PC space that separates low and high LSTM score cases.}
\label{fig:PC13_dynamic}
\end{figure}

To see the PC score dynamics for specific active regions, we plot how PC1 and PC3 scores evolve as time went by for four different active regions that have gone through sudden transition. Results are shown in Figure \ref{fig:PC13_dynamic}. The dynamics of PC1 and PC3 score across time is shown in a series of connected arrows in the PC1-PC3 score subspace. Apart from showing how the PC1 and PC3 score change for every 12 minutes using arrows, we also indicate whether the change happened during the LSTM score sudden transition time by labelling the arrows during sudden transition in green. It can be seen from Figure \ref{fig:PC13_dynamic} that when LSTM score became very high after the transition, namely above 0.7, typically the active regions will have both high PC1 and PC3, as indicated by the red arrows on the upper right corner of each graph. And during the transition time, where the LSTM score began to jump from lower than 0.3 to higher than 0.7, the trajectory of the PC1 and PC3 score escape the area around the origin and region with very negative PC3, and travel towards the region where PC1 score is above 20 and PC3 is above 5. Only $0.25\%$ of the points in Figure \ref{fig:pc13_score} are in this region, so the PC scores become the outliers of all samples in the PC score subspace after sudden transitions.

\begin{figure}[htb]
\centering
\includegraphics[width = 0.9\textwidth]{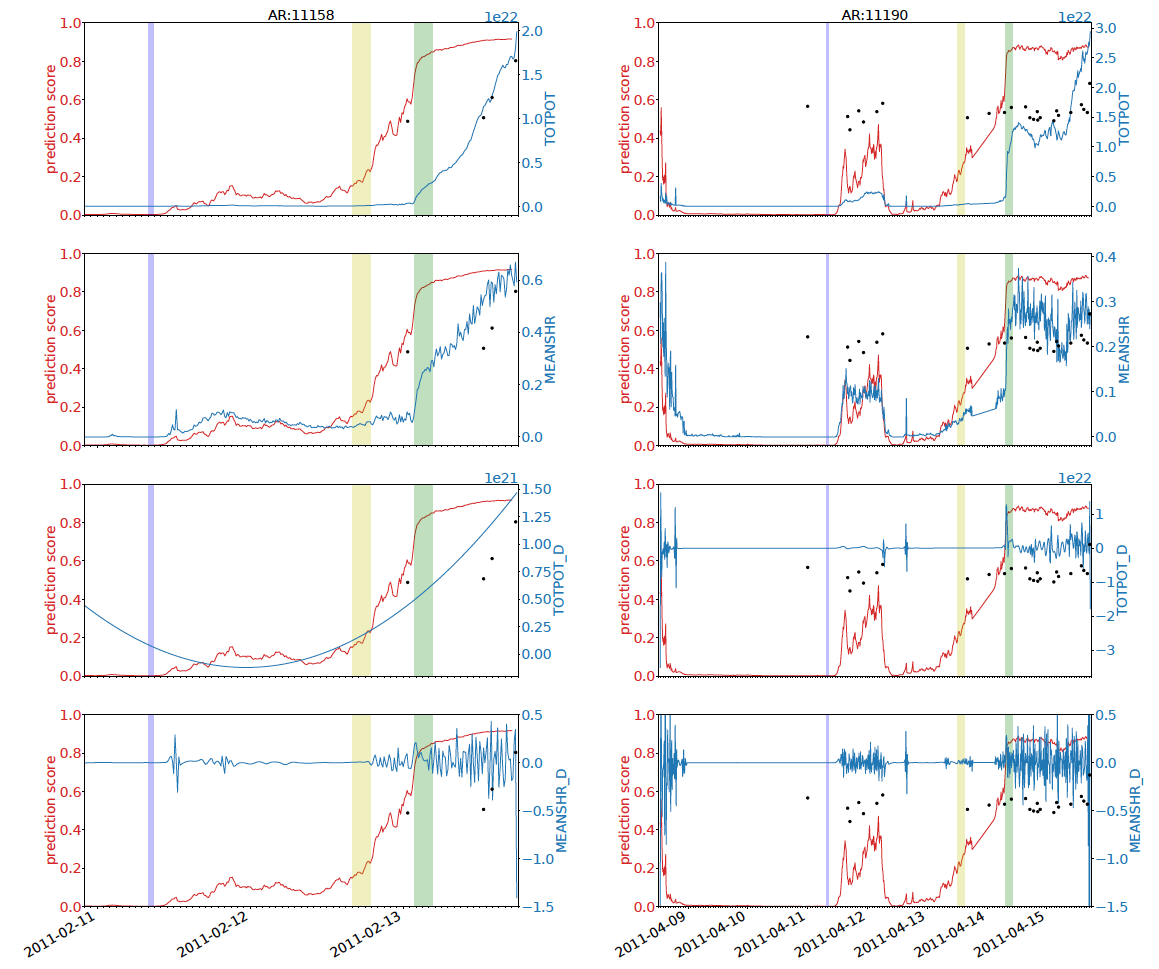}
\caption{The original SHARP parameters TOTPOT, MEANSHR and their first-order time derivatives of AR 11158 and 11190, prior to their first M flares. The blue band shows the 1-hour time-series chosen as the baseline for DTW distance calculation. The yellow and green band are a 3-hour time-series before and after the sudden transition of LSTM scores. All formats follow that in Figure \ref{fig:exampleofar}. We see that compared to the time before transition and time at the baseline, the TOTPOT and MEANSHR time-series show more increasing trends after the sudden transition. And their first-order time derivative show more volatility when compared to the time with low prediction scores. These are the driving forces behind the sudden transitions.}
\label{fig:2AR_original_highlight}
\end{figure}

To illustrate the process of sudden transitions with the original SHARP parameters instead of the DTW distance features, in Figure \ref{fig:2AR_original_highlight}, we show the original TOTPOT, MEANSHR, TOTPOT\_D, MEANSHR\_D of AR 11158 and 11190, prior to their first M flares. The blue band highlights the 1-hour time-series at the baseline point. The yellow and the green bands highlights a 3-hour time-series before and after the sudden transition of LSTM scores, featuring consistently low ($<$0.3) and high ($>$0.7) LSTM score. It can be seen in Figure \ref{fig:2AR_original_highlight} that the time-series before sudden transition is not very different from the baseline in all four quantities. But when LSTM score is very high, there is a sudden transition of TOTPOT and MEANSHR inside the green band for both active regions. In the meantime, the derivative of TOTPOT is becoming very large, and the derivative of MEANSHR has become extremely noisy. In general, when LSTM scores are higher, the main trends of SHARP parameters become very different from the baseline, and the signals of flare eruption has emerged. Additionally, the derivatives are becoming different from the baseline as well, with derivative of TOTPOT becomes higher, and derivative of MEANSHR becomes noisier.

In the case study of AR 11190 above, we see that prior to the sudden transition time, there is another time (around 2011-04-12) where the LSTM score rises above zero but went back to zero before becoming very large. During this period, the TOTPOT only rises up mildly, but the MEANSHR has become significantly different from zero. One can see that there are several flares of intermediate intensities happening during this time. Indeed, such a phenomenon is very common when weak flares (B/C flares) are upcoming. The TOTPOT is remains largely unchanged during these flares and only rises up before the last strong M/X flares. The MEANSHR, however, is more sensitive and can be fairly large even before these weaker flares. Such a sensitivity to weak flares sets the MEANSHR, and many other SHARP parameters about magnetic field gradients such as the MEANGBT, MEANGBH and MEANGBZ, apart from the SHARP parameters that are more insensitive to weaker flares, such as the USFLUX, ABSNJZH, SAVNCPP and TOTPOT. 

In fact, PC3 sets these two groups of features apart. It is giving all SHARP parameters such as TOTPOT, TOTUSJH, SAVNCPP, USFLUX positive loadings but near zero loading for MEANSHR, MEANGAM and MEANGBH. Since MEANSHR, MEANGAM and MEANGBH can become high even before some weak flares, they are worse at distinguishing LSTM inputs before a B/C flare and an M/X flare than other SHARP parameters such as TOTPOT. If we call the insensitive features such as TOTPOT as \textbf{signals}, and all the sensitive ones such as MEANSHR\_D as \textbf{noise}, then PC1 is simply the signals plus noise, and PC3 is signal minus noise. The LSTM score will be high only when the signal features are very large. If we only observe large noisy features, it can be the case that weak flares are happening. Such a SHARP parameter taxonomy echoes the finding on variable importance in \cite{Yang2019} where the accuracy of LSTM classification drops only a little if it is trained solely with TOTUSJH or TOTPOT, but drops significantly when trained solely with MEANSHR or MEANGAM.

The LSTM model cannot extract signals from noises perfectly. Because the noise features become more intense 
together with the signal features before an M/X flare. But before weak flares, it is more likely that we only see some noise features rising up. Thus LSTM cannot tell which type of flare is going to happen under the cases when MEANSHR is both high and volatile, unless it sees some significant TOTPOT as well. As one can see from the case study of AR 11190 in Figure \ref{fig:2AR_original_highlight}, a significantly non-zero MEANSHR still drives the LSTM score to nearly 0.4 during 2011-04-12 for AR 11190. Luckily, the LSTM is able to pick up the signals for strong flares. As the TOTPOT rises up for AR 11190, the LSTM score exceeds the upper bound at 0.4 and soars to nearly unity.

To summarize, some SHARP parameters have very different behaviors before weak flares and strong flares, such as the TOTPOT and all others with positive loadings in PC3. They are the true signals of strong flares picked up by LSTM model. Other SHARP parameters, such as the MEANSHR, have smaller contrasts before weak and strong flares. Their derivatives have negative loadings in PC3 and the noise they introduce might confuse the LSTM model. The analysis in this section is still very qualitative. In the next subsection, we will show the exact threshold of PC scores and SHARP parameters that can split the low and high LSTM score cases. 

\subsection{Tree-based analysis on LSTM results (good title)}\label{section:tree}

In this subsection, we quantify the exact threshold of PC scores and DTW features that separates low and high LSTM score cases with pruned classification and regression tree (CART). If an LSTM input surpasses the threshold, we can expect that the LSTM score is more likely to be high, as suggested by the pattern in (b) of Figure \ref{fig:DTW_Explained}.

Classification and regression tree (CART) is a non-parametric statistics model that partitions the feature space into several disjoint regions, and fits a model inside each region. CART uses a tree-like structure consisting of decision nodes and leaves. Each decision node has one of the input features and an optimized split value. Any input data that has the feature above or below the split value would be dispatched to different sub-trees until a leaf node is reached. At the leaf node, CART returns a categorical (for classification models) or continuous (for regression models) fitted value. For a more comprehensive background of the CART model, we recommend our readers to \cite[Chapter~9.2]{ESL}. CART is known to have low bias but over fits the data most of the time. To overcome the over fitting issue of CART, we pruned the fitted tree to have maximum depth of 3. We choose 3 as the maximum depth so as to keep a balance between classification accuracy and visualization simplicity.

Given the PCA model fitted in the previous section, we readily calculate the PC scores for the 36 DTW distance features. We only keep the top 10 PC scores for further analysis. As a result, for each time point on every leave-one-out score path, we have an LSTM prediction score associated with a 10-dimensional PC score vector. In total, we have 141,350 such PC score-LSTM score pairs. For each LSTM score, we label it as 1 if it is above 0.7, and label it as 0 if it is below 0.3, and drop it otherwise. Such a procedure leaves us 125,955 pairs of 10-dimensional PC scores and associated 0/1 class label. This is exactly the number of points in Figure \ref{fig:pc13_score}. We use the PC scores as input and fit a CART model to classify the 0/1 class label using these 125,955 pairs of data. One can think of this classification problem as finding a decision boundary to separate the red and blue points in Figure \ref{fig:pc13_score} in a 10-dimensional PC score space.

\begin{figure}[htb]
    \centering
    \includegraphics[width = 0.9\textwidth]{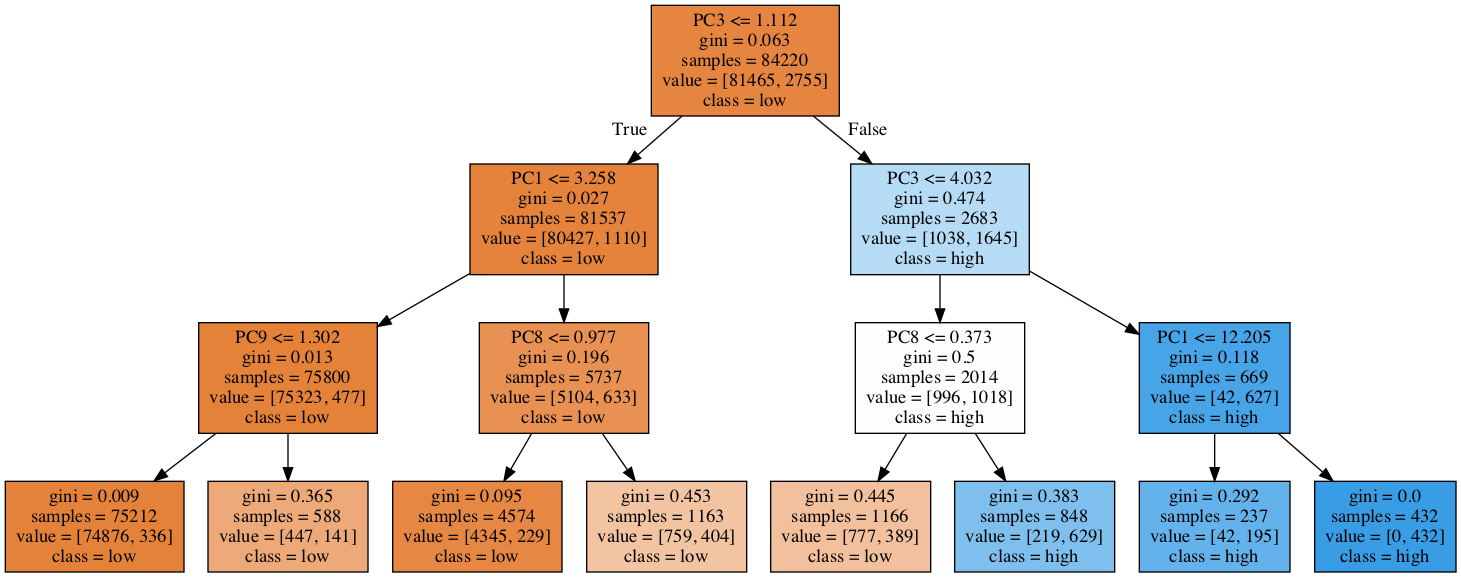}
    \caption{Classification Tree fitted using PC scores to classify low and high LSTM score cases. In each rectangle box, there is a feature and a split value in the first line, such as $\mathrm{PC3}<=1.112$ in the first box. It means that for any 10-dimensional PC score vector, if this condition is satisfied, it will be passed to the left branch, other wise it will be passed to the right branch. ``Samples" indicates how many training samples are waiting to be split at the current node. ``Value" shows the number of training samples of each class at the current node. ``Class" is the majority class of the training samples at the current node. Nodes in brown and blue are the nodes where the majority class of the samples is low and high LSTM score cases. The darkness of the colors represent the purity of the node. One can see that after thresholding all training samples with PC3 score at 1.112, the left sub-tree mainly consists of low LSTM score cases and the right sub-tree mainly consists of high LSTM score cases. The PC3 score is a very good variable to set a threshold.}
    \label{fig:CART_PCAscore}
\end{figure}

To construct a training set and a test set for the CART model, we split all samples of dataset $\mathcal{W}$ based on the active region where the samples come from. The training-test split makes sure that all pairs of PC score and 0/1 class label from the same HARP region only appear in either the training or test set. Eventually, we obtain a training set of 84,220 samples including 2,755 high LSTM score cases and a test set of 41,735 samples, including 2,419 high LSTM score cases. Figure \ref{fig:CART_PCAscore} shows the result of the CART with maximum depth limited to 3. 

The visualization of the classification tree follows a tree structure, with each rectangle box containing the information about the feature used at the node, the split value, the impurity of the node (Gini-index), the number of samples of each class remaining to be classified at the node and the majority of the class of samples at the current node. The top of the decision tree is the root node, and it is the first ``if-else statement'' of the whole tree. We can see that the root node splits the PC3 score at 1.112. All samples with PC3 score below this value are passed to the sub-tree on the left for further classification while all others are passed to the sub-tree on the right. We can see that after just one thresholding of the PC3 score, we have already obtained two sub-trees where the majority of the samples are low and high LSTM scores cases, as indicated by the color on the left and right branches after the root node splitting. The test set prediction accuracy of the classification tree is $89.3\%$.

\begin{figure}[htb]
    \centering
    \includegraphics[width = 0.9\textwidth]{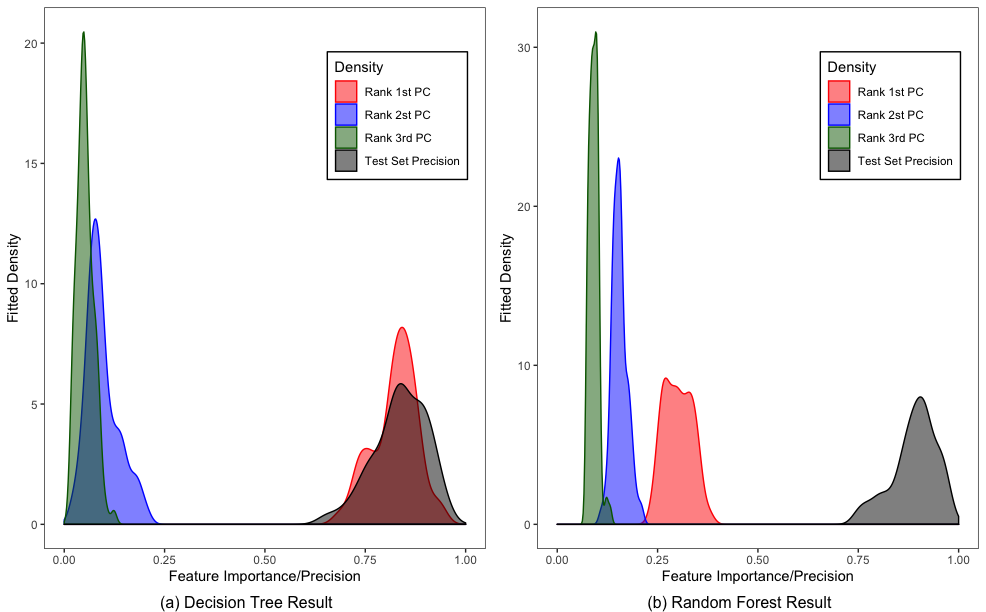}
    \caption{Results of Classification and regression tree (CART) and Random Forests, each with 100 iterations, with each iteration having a different training set and test set. In each iteration, all PC scores are ranked according to their feature importance, and we keep record of the feature importance of the top 3 PCs. Distributions of test set accuracy, feature importance of PC ranks 1st, 2nd and 3rd in each iteration's feature importance ranking are plotted. In both models, the Rank 1st PC is PC3 in all iterations. PC1 is the Rank 2nd PC $62\%$ of the time in CART and $99\%$ of the time in random forests.}%
    \label{fig:DT_RF_result}%
\end{figure}

Since our training-test splitting procedure has some randomness regarding which active regions are put in the test set, we may have different variations of the fitted CART tree if we have a slightly different training and test set. To evaluate whether the PC3 score appears consistently at the root node and plays a key role in the thresholding, we rerun the CART with maximum depth restricted to 3 for 100 iterations with different training and test sets, and calculate the feature importance of each of the 10 PC scores in each iteration. Feature importance can be interpreted as the contribution of a feature to the classification. In each iteration, we rank all PCs by their feature importance and keep records of the top 3 PCs. It turns out that PC3 is always the most important feature, PC1 is the second most important feature in 62 iterations. The left panel of Figure \ref{fig:DT_RF_result} shows the distribution of the test set prediction accuracy and the feature importance of the PC that ranks 1st, 2nd, 3rd in the feature importance ranking of 100 iterations.

The PC3 is consistently the ``Rank 1st PC" in terms of feature ranking in the 100 iterations, but the feature importance of PC3 scores and others vary a great deal as does the test set precision, which ranged from less than 0.6 to nearly 1.0. To obtain more robust results, we use random forests to do the classification of low and high LSTM scores with the PC scores. We recommend our readers to \cite[Chapter~15]{ESL} for details of random forests method. With the same training-test set splitting procedure, we run a random forest with 50 trees, with each tree being a classification tree using up to 2 randomly selected PC scores. The classification result of the random forests is the majority of the results returned by the 50 trees. Similarly, we run the random forests with 100 iterations, and in each iteration, we collect the feature importance of every PC. In the right panel of Figure \ref{fig:DT_RF_result}, we plot the distribution of the test set classification accuracy, and the feature importance of PC ranks 1st, 2nd and 3rd based on the random forests results. In the fitted random forest model, PC3 is still the most important feature in all iterations, and PC1 ranks 2nd in 99 out of 100 iterations.

\begin{figure}[htb]
    \centering
    \includegraphics[width = 0.9\textwidth]{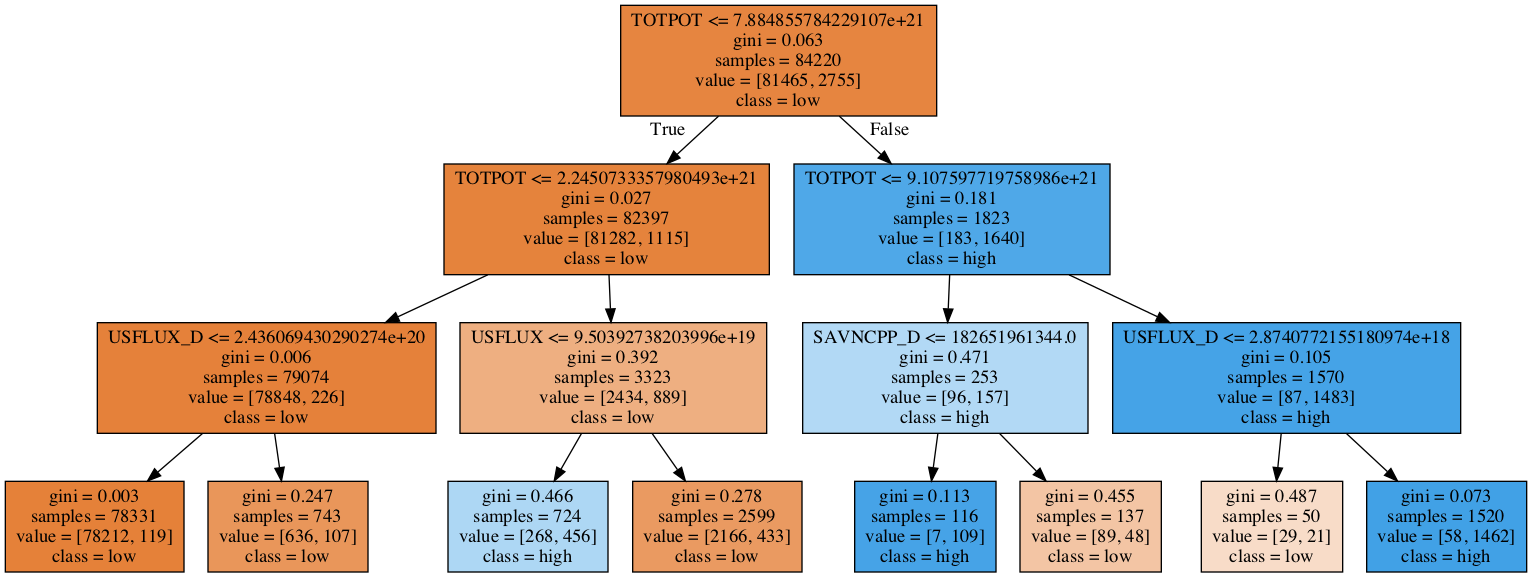}
    \caption{Classification Tree fitted using DTW distance features with maximum depth of 3. Decision nodes with the majority of the samples being the low LSTM scores are colored brown, and nodes with the majority of the samples being the high LSTM scores are colored blue. The darkness of the colors represent the purity of the node. All formats follow that of Figure \ref{fig:CART_PCAscore}. One can see that TOTPOT appears not only in the root node, but also in the root node of the two sub-trees.}
    \label{fig:DT_Original}
\end{figure}

All previous tree-based methods are applied to PC scores. The results of both tree-based methods show that the PC3 score consistently stands out as the most important feature compared to all other principal components. Simply speaking, PC3 contains the information of the signals that can predict strong flare eruption. In order to directly evaluate the association with the LSTM score for each SHARP parameter, we rerun the pruned CART model on the 36 DTW distance features instead of 10 PC scores. Figure \ref{fig:DT_Original} shows the result on the CART fitted on the DTW distance features with the same training set and test set as the ones used in generating Figure \ref{fig:CART_PCAscore}. Now, we find that by using the total free energy density (TOTPOT) as the root node feature and splitting at $7.9\times 10^{21}$ erg $\cdot \mathrm{cm}^{-1}$ results in two sub-trees with a clear pattern. The left hand side sub-tree mainly consists of low LSTM score cases and the sub-tree on the right mainly consists of high LSTM score cases. The decision tree has a test set precision at $91.2\%$.

Upon further examination, one can see that the second layer of the tree also uses TOTPOT as the decision node feature. And most of the training set cases with high LSTM scores have TOTPOT's DTW distance feature above $2.25\times 10^{21}$ erg $\cdot\mathrm{cm}^{-1}$. If one assumes that at baseline time, there is no polarity inversion line and the TOTPOT is $0$ in all 5 frames, then the threshold of TOTPOT for high LSTM prediction score is $4.5\times 10^{20}$ erg $\cdot\mathrm{cm}^{-1}$.Note that the threshold should be interpreted as a PIL mask weighted parameter. 

One caveat on interpreting the feature importance of TOTPOT in the decision tree is that many features that are highly correlated with TOTPOT could have similar feature importance as that of TOTPOT if TOTPOT is removed from the feature list. Thus in general, one can conclude that the block of variables that are highly correlated with total free energy density can be used to establish some threshold values to distinguish cases with low and high LSTM scores. Once again, the decision tree analysis confirms our previous statement that the signals picked up by LSTM model to predict strong solar flares are the SHARP parameters that are highly correlated with total free energy density, including TOTUSJH, MEANJZH, MEANALP, TOTPOT, MEANPOT, TOTUSJZ, ABSNJZH, SAVNCPP, and USFLUX.

In summary, our tree-based analysis on DTW distance features shows that when the 1-hour total free energy density time-series of an LSTM input, or the 1-hour time-series of other highly correlated SHARP parameters, becomes extremely different from the baseline 1-hour time-series of the HARP region, it is associated with high LSTM scores. Our results based on the PC scores demonstrate that there are the two groups of variables, one with positive loadings in PC3 (signals), the other with negative loadings in PC3 (noise), that play a key role out of all features in distinguishing between low and high LSTM score cases, thus weak and strong first flares.

\section{Discussion and Conclusion}\label{Conclusion}
We presents interpretations of predictions made by Long-Short-Term-Memory (LSTM) models on first flares of active regions. We use data coming from Space-weather HMI Active Region Patches (SHARPs) where parameters are calculated along the polarity inversion line (PIL). Specifically, we train the LSTM model to distinguish first B flares against first M/X flares; and generate a prediction score path for 402 first flares coming from 369 active regions. Among these events there are 35 active regions whose first M/X flare prediction score path shows a sudden transition to high probability. On average, the sudden transition to high probability is completed 48 hours before the flare. The rapid transition phenomena provide a unique opportunity to interpret the LSTM model (a black-box machine learning algorithm) and determine which physical features represented by the SHARP parameters drive the flare prediction. 

To interpret the LSTM predictions, we apply a two-step method to project the 2d matrix of inputs of LSTM into a low dimensional space. In the first step, we utilize dynamic time warping (DTW) to measure the temporal dimension similarity of any two LSTM inputs. By picking a specific LSTM input, and the baseline of the LSTM input as a benchmark, we manage to collapse the information of LSTM input matrices into a low dimensional DTW distance vector. We further reduce the dimensions in the second step by PCA. Finally, we construct a feature, the PC3, that linearly combines the similarity metric of each SHARP parameter and their derivatives. By examination of active regions experiencing a sudden transition of LSTM prediction scores with case studies, together with tree-based classification, we find that the PC3 can well distinguish LSTM inputs that give low and high prediction scores. A careful inspection of the two representative features, total free energy density (TOTPOT) and mean shear angle (MEANSHR), shows that the LSTM identifies strong flare eruptions out of a few features that carry strong signals, which frequently increase before the strong flare occurs.  The interpretation of LSTM highlights that TOTPOT and those highly correlated SHARP parameters contain the key indicators of strong flare eruption. However, the SHARP parameters highly correlated with the MEANSHR can introduce noises into the LSTM model in the sense that they can be more volatile not only during the strong flare time, but also weak flare time. In contrast, TOTPOT energy measurement is weighted by the square of the field strength, which explains its much greater stability and its greater significance for flare prediction.

The interpretation of the LSTM results allows us to identify key photospheric signatures of the energy buildup leading to flares. To give some idea of the complexity of behavior, we illustrate two representative examples where the probability of a large (M/X-class) flare rises in tandem with the buildup of total free energy (TOTPOT) and mean magnetic shear (MEANSHR) relative to pre-event baselines. In the case of AR 12017, the probability peaks with a rapid rise in free energy (and DTW) that levels off near the flare-time level.  Here, the LSTM model near-certain prediction of a large flare corresponds with levels of free energy that implies it is only a matter of time before a major flare occurs. In other words, it is very unlikely that the system can relax without the occurrence of a large flare.  In the case of AR 12381, the probability of a major flare becomes high while the total free energy is increasing but still well below the flare-time level. This example shows that the model can identify a trend of increasing free energy that is likely to continue to culminate in a flare.  The contrast in examples shows that there are distinct evolutionary paths leading to flares, each of which has unique baseline to gauge the necessary energy requirements. We find a threshold for the magnitude of total free energy density that produces a high probability of a flare at around $4.5 \times 10^{20}$ erg $\cdot\mathrm{cm}^{-1}$. 

Free energy is a necessity for solar eruptions, and fixed thresholds for initiation have been proposed \citep{Moore2012, Kostas2012, Schmieder2018, Vasantharaju2018, Wai-Leong2019}. The LSTM identifies the free energy and mean magnetic shear as critical variables for predicting flares.  This result reflects the extraordinary complexity and varying structure of ARs, that are governed by common physical processes.  The nature of those processes are elucidated by the role of the mean magnetic shear in the flare predictions.  The mean shear is a measure of the non-potential nature of the magnetic field and in these circumstances indicates that the photospheric magnetic field is oriented away from the direction of the potential field. For these PIL-confined SHARP data, this implies the magnetic field is oriented  more nearly parallel to the polarity inversion line (in contrast to a potential field, which would be perpendicular to the PIL). The sheared magnetic configuration is nearly universally observed in association with solar eruptions and our work confirms its essential role in large flares. Several leading theories offer different explanations for this magnetic shear such as sunspot collisions \citep{Chintzoglou2013, Fang2015, Chintzoglou2019}, sunspot rotation \citep{Kazachenko2009, Jiang2012, Torok2013}, super granule rotation \citep{Antiochos2013} and magnetic flux emergence \citep{Manchester2001, Fan2001, Manchester2004, Manchester2007, Archontis2008,Fang2012a, Fang2012b}. In a followup paper, we will analyze the HMI vector magnetograph data for these 35 active regions and determine in detail how the the magnetic shear developed, and determine the physical processes responsible for the buildup of energy leading to the first M/X-class flares.

Our post-hoc analysis has provided extra insights about solar flare predictions with machine learning model. Following the discussions in \cite{Yang2019}, we advance our understanding about machine learning predictions in the following aspects:

\begin{itemize}
    \item We use SHARP parameters calculated along the polarity inversion line instead of the ones calculated from the full HMI images to train LSTM model.
    
    \item We identify 35 active regions with sudden transitions of LSTM scores and have done case studies only on these selected active regions with the sharpest prediction score contrast.
    
    \item We propose a dimension-reduction technique based on dynamic time warping (DTW) and principal component analysis (PCA) to summarize the information contained in matrix-shaped LSTM inputs. The low-dimensional representation of LSTM inputs shows some very interpretable learning patterns of LSTM model.
    
    \item The constructed features in the low-dimensional space still have very good interpretability, and we show that the key feature contains signals from a subset of SHARP parameters and noises from others.
    
    \item Specifically, we show that SHARP parameters that are highly correlated with total free energy density are the important signals for strong flare eruption learnt by the LSTM model. On the contrary, SHARP parameters whose derivatives are highly correlated with the derivative of mean shear angle introduce noises to the flare eruption pattern learnt by the LSTM model. 
    
    \item A threshold for total free energy density along the polarity inversion line at $4.5 \times 10^{20}$ erg $\cdot \mathrm{cm}^{-1}$ is given, surpassing which indicates that a strong flare is about to happen.
\end{itemize}

\newpage

\appendix
\section{Summary of Sudden Transition Cases}

\renewcommand{\arraystretch}{0.85}
\begin{table}[H]
\centering
\begin{tabular}{|c|c|c|c|}
\hline 
AR Number & Prior Transition Time & Post Transition Time & Flare Time \\\hline
11410 & 2012-02-04 14:00:00 & 2012-02-06 02:48:00 & 2012-02-06 20:00:00\\\hline
11560&2012-08-30 10:00:00&2012-08-30 14:12:00&2012-09-06 04:13:00\\\hline
11613&2012-11-12 03:00:00&2012-11-12 04:12:00&2012-11-12 23:28:00\\\hline
11618&2012-11-19 18:00:00&2012-11-20 09:00:00&2012-11-20 19:28:00\\\hline
11718&2013-04-09 03:12:00&2013-04-11 05:24:00&2013-04-12 20:38:00\\\hline
11726&2013-04-20 00:36:00&2013-04-20 06:00:00&2013-04-22 10:29:00\\\hline
11762&2013-06-01 16:36:00&2013-06-02 21:12:00&2013-06-05 08:57:00\\\hline
11817&2013-08-11 07:12:00&2013-08-12 00:12:00&2013-08-12 10:41:00\\\hline
11818&2013-08-13 08:48:00&2013-08-14 17:24:00&2013-08-17 18:24:00\\\hline
11861&2013-10-10 19:36:00&2013-10-11 04:12:00&2013-10-17 15:41:00\\\hline
11891&2013-11-06 21:24:00&2013-11-07 02:12:00&2013-11-08 09:28:00\\\hline
11928&2013-12-18 21:12:00&2013-12-19 05:12:00&2013-12-22 08:11:00\\\hline
11153&2011-02-08 08:24:00&2011-02-08 13:36:00&2011-02-09 01:31:00\\\hline
11968&2014-01-30 06:36:00&2014-01-30 15:36:00&2014-01-31 15:42:00\\\hline
11158&2011-02-12 18:48:00&2011-02-13 01:48:00&2011-02-13 17:38:00\\\hline
12017&2014-03-28 01:00:00&2014-03-28 03:36:00&2014-03-28 19:18:00\\\hline
11165&2011-03-05 19:48:00&2011-03-06 01:12:00&2011-03-07 07:54:00\\\hline
11169&2011-03-09 03:36:00&2011-03-09 19:24:00&2011-03-14 19:52:00\\\hline
12065&2014-05-24 01:36:00&2014-05-24 08:24:00&2014-05-24 18:35:00\\\hline
12085&2014-06-07 17:00:00&2014-06-08 17:48:00&2014-06-12 09:37:00\\\hline
12089&2014-06-11 04:00:00&2014-06-11 13:24:00&2014-06-12 20:03:00\\\hline
12182&2014-10-02 23:00:00&2014-10-03 02:48:00&2014-10-09 01:43:00\\\hline
11190&2011-04-13 14:48:00&2011-04-14 07:24:00&2011-04-15 17:12:00\\\hline
12257&2015-01-08 19:36:00&2015-01-09 09:00:00&2015-01-13 04:24:00\\\hline
12280&2015-02-04 13:48:00&2015-02-05 01:36:00&2015-02-09 23:35:00\\\hline
12360&2015-06-11 07:12:00&2015-06-11 16:36:00&2015-06-13 07:29:00\\\hline
12381&2015-07-05 12:36:00&2015-07-06 00:36:00&2015-07-06 08:44:00\\\hline
12403&2015-08-18 10:48:00&2015-08-18 11:48:00&2015-08-21 02:18:00\\\hline
12423&2015-09-27 12:48:00&2015-09-27 18:24:00&2015-09-28 03:55:00\\\hline
12422&2015-09-25 06:48:00&2015-09-26 09:12:00&2015-09-27 10:40:00\\\hline
12644&2017-04-01 13:24:00&2017-04-01 15:48:00&2017-04-01 21:48:00\\\hline
11260&2011-07-25 22:00:00&2011-07-26 03:36:00&2011-07-27 16:07:00\\\hline
11261&2011-07-28 15:00:00&2011-07-29 09:24:00&2011-07-30 02:09:00\\\hline
11263&2011-07-30 10:12:00&2011-07-31 15:36:00&2011-08-03 04:32:00\\\hline
11283&2011-09-04 03:36:00&2011-09-04 11:48:00&2011-09-06 01:50:00\\\hline
\end{tabular}
\caption{Active regions with sudden transitions of LSTM prediction scores. Each row contains the information of the NOAA active region number, the time before and after the sudden transition and the first M/X flare time. All times are in the format Year-Month-Day Hour-Minutes-Seconds. On average, the sudden transition was completed 48 hours before the flare.}\label{tab:ST}
\end{table}

%



\newpage
\acknowledgments
We thank Yang Liu from Solar Dynamics Observatory (SDO) of Stanford University, Meng Jin from Lockheed Martin and Alfred Hero from University of Michigan for very helpful comments on the data and statistical analysis of our paper. We greatly appreciate the help on the data preparation of the polarity inversion line based SHARP parameters by Jingjing Wang from the National Space Science Center in Beijing. This research is supported by NASA grant 80NSSC18K1208.


%
%

\bibliography{interpretLSTM}

%
%
%
%
%

\end{document}